\documentclass[a4paper,11pt]{article}
\pdfoutput=1 

\usepackage{jcappub} 

\usepackage{aas_macros} 

\usepackage[T1]{fontenc} 
\usepackage{latexsym}
\usepackage{amsmath}

\title{\boldmath Astrophysical Plasma Instabilities induced by Long-Range Interacting Dark Matter}

\author[a]{A. Cruz, } \author[b]{Matthew McQuinn}

\affiliation[a]{Department of Physics, University of Washington,\\Seattle, WA 98195}
\affiliation[b]{Department of Astronomy, University of Washington,\\Seattle, WA 98195}

\emailAdd{admcruz@uw.edu}
\emailAdd{mcquinn@uw.edu}

\abstract{If dark matter is millicharged or darkly charged, collective plasma processes may dominate momentum exchange over direct particle collisions.  In particular, plasma streaming instabilities can couple the momentum of the dark matter to counter-streaming baryons or other dark matter and result in the counter-streaming fluids coming to rest with each other, just as happens for baryonic collisionless shocks in astrophysical systems.  While electrostatic plasma instabilites (such as the two-stream) are highly suppressed by Landau damping when dark matter is millicharged, in the cosmological situations of interest, electromagnetic instabilities such as the Weibel can couple the momenta, assuming that the linear instability saturates in the manner typically found for baryonic plasmas. We find that the streaming of dark matter in the pre-Recombination universe is affected more strongly by direct collisions than collective processes, validating previous constraints. However, when considering unmagnetized instabilities the properties of the Bullet Cluster merger and other merging cluster systems (which show dark matter streaming through itself) are likely to be substantially altered if $[q_\chi/m_\chi] \gtrsim 10^{-4}$, where $[q_\chi/m_\chi]$ is the charge-to-mass ratio of the dark matter relative to that of the proton.  When a magnetic field is added consistent with cluster observations, the Weibel and Firehose instabilities result in sufficiently fast growth to reach saturation for $[q_\chi/m_\chi] \gtrsim 10^{-12}-10^{-11}$.  The Weibel growth rates are even faster in the case of a dark-$U(1)$ charge (because ``hot'' electrons do not damp the instability), potentially ruling out $[q_\chi/m_\chi] \gtrsim 10^{-14}$ in the Bullet Cluster system, in agreement with \citep{Lasenby2020}.  The strongest previous limits on millicharged dark matter (mDM) arise from considering the spin-down of galactic disks \citep{Stebbins2019}.  We show that plasma instabilities or tangled background magnetic fields could lead to diffusive propagation of the dark matter, weakening these spin-down limits.  Thus, plasma instabilities may place some of the most stringent constraints over much of the millicharged, and our results corroborate previous extremely stringent potential constraints on the dark-charged parameter space.}

\begin{document}
\maketitle
\flushbottom

\section{Introduction}

Multiple, orthogonal astrophysical observational probes provide evidence for the dark matter: galaxy rotation curves \citep{Sofue2001}, the cosmic microwave background (CMB) \citep{Plank2018}, gravitational lensing \citep{Massey2010}, and the matter power spectrum \citep{Primack2015}. This collection of observations indicates that dark matter (DM) interacts gravitationally, but leaves room for some other interactions. In well-motivated DM models such as weakly interacting massive particles (WIMPS), DM also interacts with the Standard Model (SM) through mediators at the weak scale.  The QCD axion, on the other hand, has a weak effective electromagnetic interaction but no electric charge. If instead DM has a small ``millicharge'' and has long-range electromagnetic (EM) interactions with the SM, there can be significant effects on astrophysical and cosmological scales. One such model being the dark photon, which gives rise to kinetic mixing and naturally produces DM with a small electromagnetic charge \citep {Liu2019}. This DM model is known in the literature as millicharged DM (mDM) and allows for long-range EM interactions between DM and the SM. 
 In this model, the particle DM has mass $m_{\chi}$ and a small electromagnetic charge $q_{\chi}$ as well as its antiparticle. While the nomenclature used in the literature is to refer to this DM model as ``millicharged", the considered range of $|q_{\chi}|$ spans many orders of magnitude.

One benchmark millicharge occurs in freeze-in models \citep{Dodelson1994, Chu2012}, where over a large range of dark matter masses a coupling with $q_\chi/e \sim 10^{-11}$ is able to produce the relic abundance of DM via decays from SM particles.  Such minuscule charge-to-mass ratios can occur in string theoretic constructions \citep{Shiu:2013wxa}. Freeze-in models have received a lot of interest as a way of producing light DM for which there are many new experimental search techniques, with $m_\chi \sim 10-1000$ keV \citep{Battaglieri2017}.  Freeze-in models with dark matter masses $m_\chi \lesssim 0.1$ MeV are ruled out by stellar cooling constraints \citep{Dvorkin2019}.  CMB and large-scale structure observations have the potential to rule out somewhat larger masses, as direct collisions transfer momentum and damp density fluctuations on small scales \citep{Dvorkin2019}.  Perhaps most excitingly, all freeze-in parameter space up to $100$ GeV may be ruled out by spin-down of the Milky Way galaxy \citep{Stebbins2019}.  We will revisit this spin-down constraint in light of our calculations.

Another possibility is that the dark matter is charged in a $U(1)$-gauged dark sector that does not interact with SM particles \citep[e.g.][]{2013arXiv1311.0029E}.\footnote{A weak mixing of this sector with the SM hypercharge is often the case for how millicharge is created.  In this case, whether we are in this limit or the millicharged limit depends on whether dark matter-baryon or dark matter-dark matter interactions are stronger.}  This possibility avoids the anomaly conditions and tight observational constraints that arise from SM hypercharge.  In a purely darkly charged model in thermal equilibrium with an associated massless dark photon, constraints arise from limits on the number of degrees of freedom at Big Bang Nucleosynthesis and at Recombination \citep{2019JCAP...10..055A}. 
Models with sectors that mirror the SM complexity provide additional constraints \citep{2010JCAP...05..021K, 2013PDU.....2..139F, 2018PhRvD..97l3018G}. Some constraints still apply if the dark photon that mediates the interaction is ultralight but not massless.  Limits on a charged dark sector (including sectors also with SM millicharge) are reviewed in \cite{2013arXiv1311.0029E}.

Many of the cosmological constraints on millicharged and dark charged particles have largely focused on direct collisions.  For the baryonic plasmas  we are most familiar with, collisions are rarely the most effective mechanism for transferring momentum.  Rather, there are collective processes that couple the momenta of counter-streaming plasmas.  This results in collisionless shocks, where the width of the shock is much narrower than the collisional mean free path. These shocks occur ubiquitously in astrophysics \citep{collisionlessshocks}.  Just like with the baryons, long range collective forces between DM particles or -- in the case of mDM -- with the baryonic matter can couple the momentum of counter-streaming flows.  This collective behavior could dominate DM momentum coupling that assume only collisions, and could improve CMB and the Bullet Cluster constraints on such coupling, such as \citep{Heikinheimo2015}.  

We consider plasma streaming instabilities that would act to couple the momentum of the DM to either the gas or to the DM itself.  Streaming instabilities have been previously considered for mDM in \cite{2015PhLB..749..236H, 2017A&A...608A.125S, LiLin2020, 2020JCAP...11..034L, 2021AIPA...11f5013X}. When DM is millicharged,  electrostatic instabilities, or instabilities with any significant longitudinal component, are highly Landau damped owing to the streaming velocity (and hence the phase velocity of the excited Langmuir waves) being not much larger than the particle velocity dispersions for the cosmological situations of interest \citep{2021AIPA...11f5013X}.  Thus, electromagnetic instabilities are the most promising when dark matter is millicharged. On the other hand, when DM has a dark-$U(1)$ charge \citep{ Lasenby2020} has shown that the growth of the electrostatic two-stream instability grows at the dark matter plasma frequency. On the other hand, 
electrostatic instabilities, like the two-stream, generally saturate by flattening the bump on the tail of the distribution.  Such saturation may occur before substantial momentum exchange \citep[e.g.][]{1997aspp.book.....T}.  Thus, we also focus on the electromagnetic instability in this case.  We consider primarily the Weibel instability \citep{Weibel1959, fried59}. 
Indeed, the Weibel instability is thought to mediate collisionless shocks in astrophysics \citep{2014PhPl...21g2301B}, saturating in a manner where the penetrating plasmas come to rest. Although this case has been considered in the literature \citep{Ackerman2009}, we expand upon previous analysis and verify the cold approximation for Weibel instability growth rate for the Bullet Cluster system. We also consider the Firehose instability, which requires an initial background magnetic field. \\

In this paper we examine the long-range collective electromagnetic interactions of mDM and dark-$U(1)$ charged DM in galaxies, clusters of galaxies, and in the CMB. Our paper is outlined as follows: 
\begin{enumerate}
    \item In Section \ref{sec:plasma_insta}, we provide a brief introduction of plasma instabilities. We  then discuss the time required to reach instability saturation, and the time-scales associated with two particle collisions, which alter the collective plasma effects and destroy the growth of instabilities.
    \item In Section \ref{sec:electorstatic} we discuss electrostatic plasma beam instabilities and show that they are likely significantly damped if DM is millicharged.
    \item This brings us to Section \ref{sec:electromagnetic}, where we discuss electromagnetic plasma instabilities. In particular, the relevant instabilities are the Weibel instability (\S \ref{sec:weibel}) and the Firehose instability (\S \ref{sec:firehose}). In the former case we consider three astrophysical cases: mDM or $U(1)$ interactions during cluster mergers, streaming DM in the Recombination-era plasma, and we briefly discuss mDM falling onto the Milky Way. 
    We calculate instability growth rates and estimate when these growth rates are fast enough to grow within the age of the system and to not be suppressed by two particle collisions.
    
    \item In Section \ref{sec:previousconstraints}, we revisit previous dark charged constraints in light of our findings. In particular, we revisit the strong mDM constraint from the spin-down of galactic disks of \cite{Stebbins2019} in the limit where DM has a mean free path to scatter off magnetic inhomogeneities that is much smaller than the size of the halo, as may happen if the aforementioned plasma instabilities are able to reach saturation.
    
    \item In Section \ref{sec:discussion} we discuss our mDM and dark-$U(1)$ parameter space constraints on the $q_{\chi}$-$m_{\chi}$ plane. Additionally, we review and discuss previous constraints set in the literature in light of our findings.  
    
    \item Finally, in Section \ref{sec:conclusion} we discuss our constraints, illustrating the new mDM and darkly charged parameter space that is excluded by our study.
\end{enumerate}

\section{Plasma Instabilities} \label{sec:plasma_insta}

Charged particles participate in long-range collective electromagnetic interactions that enhance their momentum exchange rate relative to particle-particle interactions.  Streaming instabilities, which arise when a plasma of charged particles streams through a plasma at rest, are one of the most studied manifestations of these collective forces.  For electrons and protons, the collective interactions dominate over single particle interactions and lead to momentum redistribution between the inter-penetrating plasmas on timescales much shorter than the mean collision time. Single particle collisions between the mDM and baryons goes as the square of the charge-to-mass ratio, whereas instability growth rates can be linear or even the $2/3$ power in the mDM charge-to-mass ratio relative to protons, which we denote as $[q_\chi/m_\chi]$.  This suggests that instabilities may be the dominant mode for momentum exchange in the interesting parameter space of $[q_\chi/m_\chi] \ll 1$.  

However, a plasma instability will only grow if its growth rate is faster than the rate at which two-particle collisions dissipate the waves.  For mDM, the collisions that dissipate waves the fastest are electron-electron collisions.  Electron-electron collisions damp waves excited by the instability with a rate of \citep{Huba2013}
\begin{equation}
\label{eqn:eecollisions}
\Gamma_{\rm coll}^{ee} = 0.3 {\rm ~s^{-1}} \left( \frac{n_e}{10^{3}{\rm \; cm}^{-3}} \frac{\Lambda}{30}  \right) \left(\frac{T}{5000{\rm 
\; K}}\right)^{-3/2}.
\end{equation}
If this rate is faster than the linear instability growth rate, the instability will be quenched.  As the electron-electron collision time is much shorter than the baryon-dark matter collision time for relevant mDM parameter space, we never need to consider whether the growth rate of plasma instabilities is faster than the collisional coupling rate of the dark plasma to the baryonic plasma.  For DM charged under a dark-$U(1)$, the analogous dark matter--dark matter collision time is too long to damp the instability for the small dark charges of interest.

Another timescale of relevance is the age of the system. The instability has to happen within 1\% of the age of the system to reach saturation, as $100$ e-foldings is likely needed.\footnote{To derive the rough ${\cal N} \sim 100$ e-foldings criteria, a conservative assumption is that each electromagnetic mode starts with a (minuscule) thermal amplitude of $k_b T$, where $T$ is some characteristic temperature.  The Weibel instability grows by ${\cal N}$ e-foldings until the magnetic field reaches equipartition \citep{Medvedev1999} so that $ k_{\rm max}^3 k_b T \exp[{\cal N}] \sim  B_{\rm eq}^2$, where $k_{\rm max}$ is the fasting growing mode, and $B_{\rm eq}^2 \sim n k_b T$, where $n$ is the particle number density.  This reduces to the condition that ${\cal N} = \ln(n/k_{\rm max}^3) \approx 80 -0.5 \ln (n/10^{-5}{\rm cm}^{-3}) - 1.5 \ln ([q_\chi/m_\chi]/10^{-3})$, where we have used our later result that $k_{\rm max} \sim w_{p \chi}/c$ (\S~\ref{sec:electromagnetic}).}  In the case of a successful Weibel instability, saturation corresponds to growing magnetic fields to the point at which they can deflect streaming particles on the scale of the instability.  

\begin{table}[tbp]
\centering
\begin{tabular}{c c c c c }
\hline
\hline
plasma frequency& $\omega_{\mathrm{p}j}$ = $\sqrt{4 \pi n_j q_{j}^2
/m_j}$  & Alfv\'en speed & $v_A \equiv\Sigma_j B_0/ \sqrt{4 \pi \rho_j}$ \\
Larmor frequency  & $\Omega_j \equiv q_j B_0/m_j c $ & Larmor radius & r$_{\mathrm{L},j} \equiv m_j c v_j/q_j B_0$ \\
thermal speed & $\sigma_{T, j} \equiv \sqrt{2 kT_j/m_j}$ & Debye screening length & $\lambda_{\mathrm{D}, j} \equiv \sqrt{kT_j/4\pi n_j q_j^2} $ \\
\hline
\hline
\end{tabular}
\caption{\label{tab:plasmaparams} Definitions of commonly used plasma parameters in this work, in CGS-gaussian units. }
\end{table}

\begin{figure}[h!]
    \centering
    \includegraphics[scale=0.3
    ]{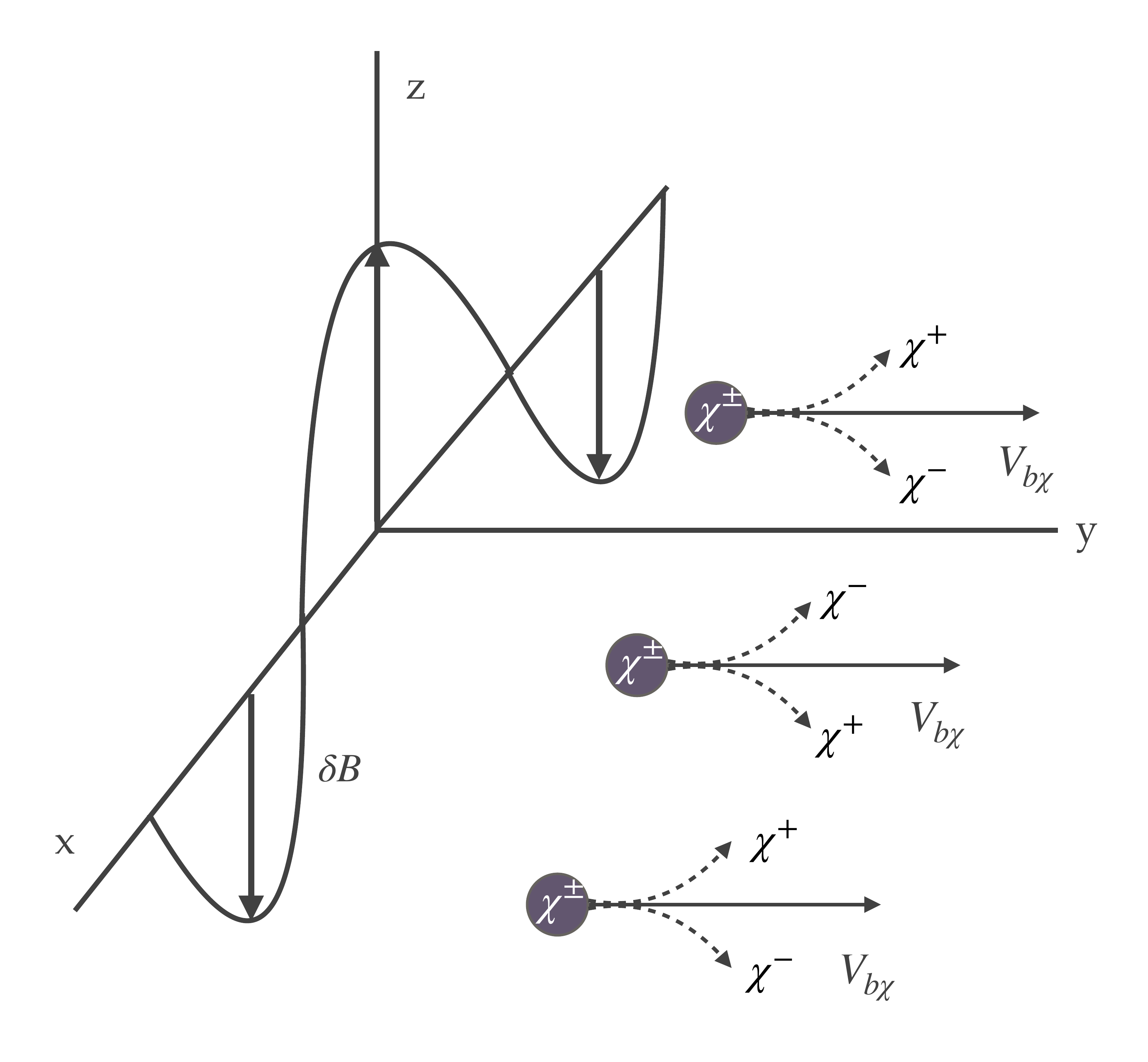}
    \caption{An illustration of the Weibel instability driven by the mDM beam in a perpendicular shock and in the absence of a background magnetic field. A magnetic fluctuation $\delta${\bf B} is assumed to oscillate in $x$ and $t$ and to point perpendicular to the direction of the mDM beam, which is in the $\hat{\bf y}$ direction. The magnetic fluctuation is thus in the $x$-$z$
    plane. A background plasma, composed of either electrons/protons or positive/negative charged DM is not shown, but the presence of the background plasma is critical for instabilities.}
    \label{fig:SNe_Dkgamma}
\end{figure}

\section{Electrostatic Plasma Instabilities} \label{sec:electorstatic}

We briefly discuss electrostatic plasma beam instabilities here and why they are likely significantly damped in the mDM case (although not necessarily in the dark-$U(1)$ case). 
Streaming can excite (electrostatic) Langmuir and ion acoustic waves that have phase velocities near the streaming velocity.  The cosmological situations of interest have particle velocity dispersions that are comparable to the phase velocity of the waves.  Thus, when DM is millicharged, the excited waves are Landau damped by the ample supply of charged particles traveling with the waves.  For example, in the two-stream instability (the most-famous-instability-of-all!), the growth rate $\gamma$ for a warm dark matter beam with velocity dispersion $\sigma_{T \chi}$ and warm background plasma with $\sigma_{T j}$ (which could be electrons, protons or dark matter -- often really all three\footnote{In the case of all three, their damping effects should be summed in the following equation.}) satisfies the proportionality \citep{1997aspp.book.....T}
\begin{equation}
    \gamma \propto \overbrace{(V_{b\chi} - \omega_r/k) \exp\left[-\frac{({\omega_r}/{k} - V_{b\chi})^2}{2  \sigma_{T\chi}^2}\right]}^{\rm growth}  
    - \overbrace{\left(\frac{\omega_{{\rm p}j}^2 \sigma_{T\chi}^3}{\omega_{{\rm p}\chi}^2\sigma_{Tj}^{3}} \frac{\omega_r}{k}\right) \exp{\left[-\frac{\omega_r^2}{2 k^2 \sigma_{Tj}^2}\right] }}^{\text{Landau damping}}
\label{eqn:electrostatic}
\end{equation}
where $\gamma$ is the imaginary component of the solution that leads to instability and $\omega_r$ is the real frequency of the instability which needs to satisfy $V_{b\chi} > \omega_r/k$ for there to be any instability, but the instability weakens considerably once $V_{b\chi} \gg \omega_r/k$ owing to the exponential factor in the first term on the right-hand side.  However, if $\omega_r/k \sim V_{b\chi}$ the dark matter needs to be very cold with $\sigma_{T\chi}^3/\sigma_{Tj}^3 \ll \omega_{{\rm p}\chi}^2/\omega_{{\rm p} j}^2 \approx [q_\chi/m_\chi]^{2}/[q_j/m_j]^{2}$ for the second term on the right-hand side, due to Landau damping, not to dominate and suppress instability, in the interesting limit for mDM damped by electrons, $[q_\chi/m_\chi]/[q_e/m_e]$ is very small. Note that velocity dispersion $\sigma_{T j}$, the plasma frequency $\omega_{{\rm p}j}$ for a given species $j$ and other relevant plasma parameters used in this work are defined in Table \ref{tab:plasmaparams}. Equation~\ref{eqn:electrostatic} is for growth in the warm limit, and does not capture the contribution from the principal value of the integral that is in the plasma dispersion function.  However, owing to $\sigma_{T\chi} \sim \sigma_{Tj}$ in the cosmological situations we consider where streaming instabilities can occur\footnote{The early universe case being the exception but suffers from a huge collision rate that damps these instabilities.}, we expect that even when this component is included, the electrostatic instabilities for mDM are highly damped. However, in the case where DM has a dark-$U(1)$ charge, $[q_\chi/m_\chi]^{2}/[q_j/m_j]^{2} = 1$ since $j=\chi$ and so electrostatic instabilities can survive Landau damping \citep{Lasenby2020}.

The conclusion that electrostatic instabilities  are significantly suppressed in the mDM case echo \cite{LiLin2020}, who considered electrostatic instabilities for the case of mDM being excited by supernovae blast waves.  The authors found that these were significantly damped, with only the electromagnetic instabilities showing growth.  Similar damping results for instabilities in which the wave vector is at an oblique angle to the beam, as the electric field along the beam is again Landau damped \citep{1997aspp.book.....T, gary_1993}, the primary mathematical difference being the parallel component of wave vector appears in the Landau damping term in an analogous growth rate equation to that of Equation~\ref{eqn:electrostatic}.

The situation is different for electromagnetic instabilities for which the electromagnetic fields are perpendicular to the wave vector, as electromagnetic waves can either have phase velocities that are much larger than the particle velocities or instead because the electromagnetic streaming instability can be purely imaginary, as is the case of the Weibel instability, which we especially focus on. 

\section{Electromagnetic Plasma Instabilities} \label{sec:electromagnetic}
In this section, we turn to the possibility that charged DM (millicharged or dark-$U(1)$) excites transverse EM waves in both the absence and presence of a background magnetic field. As in \citep{LiLin2020}, we consider only the case where wave propagation is along ${\bf B_{\rm 0}}$ (i.e. ${\bf k} \parallel{{\bf B_{\rm 0}}}$) and we restrict our analysis to two cases: the Weibel instability (where ${\bf V}_{b \chi} \perp{\bf B_{\rm 0}}$) and the Firehose instability (where ${\bf V}_{b \chi} \parallel{\bf B_{\rm 0}}$). In both limiting cases, we follow standard practices and search for purely imaginary solutions to the linear dispersion relation as expressed in \citep{gary_1993}. In either case, we search for analytic solutions when possible; when this search is stifled, we instead search for numerical solutions using the method detailed in the section below.

\subsection{Numerical Method} \label{sec:numericalmethod}
Since the linear dispersion is a complex function, we look to find purely imaginary frequency solutions where the square modulus of the linear dispersion, $D(k, \omega)$,  is zero, for a given wavenumber.  It is far from trivial to find roots or to know whether the root an algorithm returns represents the fastest growing mode.  Thus, finding solutions is a fraught exercises that depends somewhat on the algorithm.  Thus, we provide some details on the algorithm we use to find roots.  We note that, except in the Bullet cluster case with a magnetic field, we use an analytic analysis to motivate the starting point of the numerical root finder.

We minimize the square modulus of the dispersion $|D|^2$ using the Nelder-Mead algorithm, a simplex search algorithm for multidimensional unconstrained optimization of a given nonlinear function $f: \mathbb{R}^n \rightarrow \mathbb{R}$ \citep{Nelder1965} (in our case $|D|^2: \mathbb{R}^2 \rightarrow \mathbb{R}$). A simplex $S$ in $\mathbb{R}^n$ is defined as the smallest subset of $\mathbb{R}^n$ that contains each whole line segment joining any two points of $n+1$ vertices $x_0,...,x_n \in \mathbb{R}^n$; i.e. a simplex is a triangle in $\mathbb{R}^2$ and a tetrahedron in $\mathbb{R}^3$. The Nedler-Mead algorithm constructs an initial working simplex $S$ around an initial input point $x_0 \in \mathbb{R}^n$. In our application of the algorithm, the simplex is initialized around an initial frequency, $\omega_0$, chosen to be an analytic solution to the linear dispersion at hand (except for the magnetized Bullet cluster case where we do not have such a solution; see below for discussion of these solutions). At each iteration of the algorithm, the vertices of the simplex are evaluated and transformed via reflection, contraction, or expansion towards the vertex which minimizes the non-linear function $|D|^2$ most, and then a new simplex is constructed. This process is repeated until the simplex $S$ is sufficiently small, the function values are close for each vertex in the simplex, or the number of iterations exceeds a maximum value, which ever comes first.  We tried other numerical methods such as gradient-decent and Newton's method and discovered that the Nelder-Mead algorithm produced the most stable results. 

\subsection{Weibel Instability} \label{sec:weibel} 
The growth rate of the linear Weibel instability can be calculated from the electromagnetic dispersion relation.  This relation for the case where the particle distribution functions can each be modeled as Maxwellian with velocity dispersion $\sigma_{T, j}$ is \citep[see for example reference][]{gary_1993}:
\begin{equation} 
\begin{split}
        0 = D^{\pm}(k, \omega) =  c^2 k^2 - \omega^2 - \sum_{b = i^+, e^-} \omega_{p b}^2 \big(\frac{\omega}{k \sigma_{T, b}}\big) Z(\xi_{b}^{\pm}) 
        \\
        -  \sum_{s = \chi^+, \chi^-} \omega_{ps}^2 \bigg[ \bigg(\frac{\omega}{k \sigma_{T, s}}\bigg)Z(\xi_{s}^{\pm}) + f_\chi \bigg(\frac{V_{b\chi}}{\sigma_{T, s}}\bigg)^2 (1 + \xi_{s}^{\pm} Z(\xi_{s}^{\pm})) \bigg],
\end{split}
\label{eqn:fulldispersion}
\end{equation}
where $Z(\xi)$ is the plasma dispersion function (see Appendix~\ref{ap:dispersionfunction}), $\xi_{b}^{\pm} = \frac{\omega \pm \Omega_b }{k \sigma_{T, b}}$, $\xi_{s}^{\pm} = \frac{\omega \pm \Omega_s}{k \sigma_{T, \chi}}$ and $f_\chi$ is the fraction of dark matter streaming.  Sometimes we will write a solution using the notation $\omega = \omega_{r} + i \gamma$.  Note that we have chosen to account for the half of the dark matter that has like charge in the number density that appears in $\omega_{p\chi}$.  Even though the dark matter is not thermal, in dark matter halos it is often reasonably described by a Maxwellian, and in the early universe the solution does not depend on $\sigma_{T, \chi}$.  Even though we start by considering the unmagnetized limit, Equation~\ref{eqn:fulldispersion} allows for a homogeneous beam-aligned magnetic field $B_0$; we will consider the case that it is nonzero in \S~\ref{sec:withmagneticfields}. When the background magnetic field is zero then of course the cyclotron frequencies $\Omega_j=0$ for all species $j$. 

\subsection{Unmagnetized Weibel Instability} \label{sec:unmagweibel} 

We first consider the unmagnetized electromagnetic Weibel instability. The streaming motion of dark matter plasma relative to other particles can drive this instability \citep{Weibel1959, fried59}. 
Figure~\ref{fig:SNe_Dkgamma} illustrates the instability.  We consider a scenario where a fraction $f_{\chi}$ of the dark matter participates in a neutral mDM beam with equal parts $+q_{\chi}$ and $-q_{\chi}$ and with beam velocity $V_{b \chi}$ that is streaming toward the other $1- f_{\chi}$ of the dark matter plus  the baryons.  We assume the dark matter to baryon ratio is the cosmological ratio of $\Omega_d / \Omega_b$ $\approx 0.2$.
Figure \ref{fig:SNe_Dkgamma} illustrates that a small fluctuation in the magnetic field $\delta B$ perpendicular to the dark matter ``beam'' results in the plus and minus charges to separate into filaments (indeed, Weibel is often called the filamentation instability).  This separation in turn enhances the amplitude of the magnetic field fluctuation, causing the filamentation and magnetic field fluctuations to grow.  The classical Weibel solution is purely growing, i.e. Re$[\omega] =0$, such that it is not suppressed by Landau damping.  This instability is found to saturate once the Larmor radius of the dark matter is comparable to the length scale of the instability \citep{Medvedev1999}, resulting in the dark matter no longer being able to stream relative to the background plasma.  Such momentum coupling cannot happen for dark matter streaming relative to the baryons in several cosmological scenarios (such as the pre-Recombination plasma or in the Bullet Cluster), potentially allowing us to put limits on the dark matter's charge. In order to find solutions to Equation~\ref{eqn:fulldispersion}, we consider the limiting behavior of $Z(\xi_j)$ (detailed in Appendix \ref{ap:dispersionfunction}) for different particle species $j$, where $\Omega_j = 0$ and thus $\xi_{j}^{\pm} = \xi_{j} = \frac{\omega}{k \sigma_{T, j}}$. Note that $\xi_j$ indicates whether the instability is cold/warm, meaning it grows faster/slower than the thermal motion of particles across the instability scale.

\paragraph{Warm limit, where $\xi_j \ll 1$ for all participating species $j$:} 
Let us first treat the unmagnetized Weibel instability in the warm limit where the $\xi_j \ll 1$ such that $Z(\xi_j) = i\sqrt{\pi} + {\cal O}(\xi_j)$.  (Expansions of $Z(\xi_j)$ are given in Appendix~\ref{ap:dispersionfunction}.)  Only keeping the dominant $j$ particle  and the dark matter steaming term Equation~\ref{eqn:fulldispersion} reduces to
\begin{equation}
0=c^2k^2 - \omega^2 - i \sqrt{\pi} \omega_{\mathrm{p}j}^2 \bigg(\frac{\omega}{k \sigma_{T, j}}\bigg) - f_\chi  \omega_{\mathrm{p}\chi}^2 \bigg(\frac{V_{b\chi}}{\sigma_{T, \chi}}\bigg)^2,
\label{eqn:warm}
\end{equation}
where the subscript $j$ of the participating species are summed over. In most applications, $j= e^-$, but $j = \chi^+ + \chi^-$ for a dark charge, with $n_{j} = n_{\chi} = 2 n_{\chi^{\pm}}$ such that $\omega_{\mathrm{p}j}^2 = \omega_{\mathrm{p}\chi}^2 = 2\omega_{\mathrm{p}\chi^\pm}^2$.

Equation~\ref{eqn:warm} has the solutions
\begin{equation}
\omega = -\frac{i \sqrt{\pi}}{2}  \bigg(\frac{\omega_{\mathrm{p}j}^2}{k \sigma_{T, j}}\bigg) \pm \frac{1}{2}\sqrt{-\pi  \left(\frac{ \omega_{\mathrm{p}j}^2}{k \sigma_{T, j}} \right)^2 - 4\left( \omega_{\mathrm{p}\chi}^2  f_\chi \bigg(\frac{V_{b\chi}}{\sigma_{T, \chi}}\bigg)^2 -c^2 k^2 \right)}.
\label{eqn:omegaweibel}
\end{equation}

Keeping the Weibel mode (${\rm Re}[\omega]\approx 0$, $\rm{Im}[\omega] > 0$) in the limit that the second term under the radical is small yields
\begin{equation}
    \gamma^{\rm W} = \frac{1}{\sqrt{\pi}}\left(\frac{k \sigma_{T, j}}{ \omega_{\mathrm{p}j}^2} \right)\omega_{\mathrm{p}\chi}^2  f_\chi \bigg(\frac{V_{b\chi}}{\sigma_{T, \chi}}\bigg)^2,
\label{eqn:gammawarm}
\end{equation}
where $\gamma^{\rm W} = {\rm Im}[\omega]$ and the superscript ``W'' stands for ``warm'' and we have dropped the $c^2k^2$ term, valid when  $c k \lesssim c k_{\rm max} \equiv  \omega_{\mathrm{p}\chi}  (f_\chi)^{1/2} ({V_{b\chi}}/{\sigma_{T, \chi}})$.  Numerically we find that instability goes away when not in this limit.  Evaluating at $k = k_{\rm max}$ leads to an estimate for the maximum growth rate assuming $j=e^-$ of
\begin{eqnarray}
    \gamma_{\rm max}^{\rm W} &=& \frac{1}{\sqrt{\pi}}\left(\frac{\omega_{\mathrm{p}\chi}^3 f_\chi^{3/2}   \sigma_{T, j}}{ c\omega_{\mathrm{p}j}^2} \right)\bigg(\frac{V_{b\chi}}{\sigma_{T, \chi}}\bigg)^3,\label{eqn:gammawarmmaxSM} \\
        &\approx& 8.3\times10^{-14} s^{-1} \left(\frac{[q_\chi/m_\chi]}{10^{-4}} \right)^{3} \left( \frac{n_e}{10^{-3}{\rm cm}^{-3}} \right)^{1/2} \left(\frac{f_\chi}{0.5} \right)^{3/2} \bigg(\frac{V_{b\chi}}{\sigma_{T,\chi}}\bigg)^3, \nonumber
\end{eqnarray}    
where we have evaluated at values most applicable for our mDM Bullet Cluster case (and the fiducial parameters are given in Table \ref{tab:values}), and we remind the reader that $[q_\chi/m_\chi]$ is the dark charge-to-mass ratio with respect to the proton's.\footnote{After the fact, one can verify that the approximations that led to our expression for $\gamma_{\rm max}^{\rm W}$ hold.  For example, our approximation  $Z(\xi) \approx i\sqrt{\pi}$ only holds for the branch where ${\rm Im}[\xi] < 1/{\rm Re}[\xi]$ (see Appendix~\ref{ap:dispersionfunction}).  While the growth we found is purely imaginary solution so that ${\rm Re}[\xi] = 0$ clearly holds, keeping the next order in the expansion for $Z(\xi)$ results in ${\rm Re}[\xi]/{\rm Im}[\xi] \sim \xi_j$, where for our application with $j=e$.  Thus since ${\rm Re}[\xi] \ll {\rm Im}[\xi] \ll 1$ for our solution, we are safely in the desired limit.} 
The growth in this warm limit scales with the dark matter charge-to-mass to the cubic power, unlike the limits that follow where the growth scales linearly.

\paragraph{Warm-cold limit, where electrons have $\xi_e \ll 1$ and the dark matter $\xi_\chi \gg 1$:}
There is a second limit that we find is applicable in the pre-Recombination plasma --  when the dark matter is cold and other species are warm.  In order to be cold in the unmagnetized case $\omega > k \sigma_{T, \chi}$, which using our expressions for $k_{\rm max}$ and $\gamma_{\rm max}$ reduces to the condition at the wavenumber of maximum growth that 
\begin{equation}
[q_\chi/m_\chi] \gtrsim (\sigma_{T, \chi}/V_{b\chi})^{2} (\sigma_{T, \chi}/\sigma_{T, j}) f_S^{-1/2}.
\label{eqn:coldcondition}
\end{equation}
This inequality is satisfied in the early universe (before dark matter halo formation) as the dark matter velocity dispersion, $\sigma_{T, \chi}$ is essentially zero.  In this case, Equation \ref{eqn:fulldispersion} reduces to 
\begin{equation}
    c^2k^2 - \omega^2 - i \sqrt{\pi} \omega_{\mathrm{p}j}^2 \bigg(\frac{\omega}{k \sigma_{T, j}}\bigg) + f_\chi\omega_{\mathrm{p}\chi}^2 \left( \frac{V_{b\chi} k}{\omega} \right)^2 = 0,
\end{equation}
and the solution to the quartic equation in the limit that $c^2k^2$ and $\omega^2$ terms are smaller than the two final terms is
\begin{equation}
\gamma^{\rm WC} \approx \left(\frac{f_\chi}{\sqrt{\pi}} \frac{\omega_{p\chi}^2}{\omega_{pj}^2} (k \sigma_{T, j})  (k V_{b\chi})^2 \right)^{1/3},
\label{eqn:gammawarmcold}
\end{equation}
where the solution is valid for $c k < c k_{\rm max}' \equiv  \omega_{pj} (\sqrt{\pi} f_\chi  (\omega_{p\chi}/ \omega_{pj})^2 V_{b\chi}^2/\sigma_{T, j}^2)^{1/6}$.  Plugging in $k_{\rm max}^{\
\rm WC}$ into Equation~\ref{eqn:gammawarmcold} yields
\begin{eqnarray} \label{eqn:gammawarmcoldmax}
\gamma^{\rm WC}_{\rm max} &\approx& \omega_{pj} \pi^{-1/6} \left(f_\chi \right)^{1/2} \left(\frac{\omega_{p\chi}}{\omega_{pj}}\right) \left(\frac{V_{b\chi}}{c} \right),\\
 &=& 1.2\times10^{-1} {\rm s}^{-1}~ f_\chi^{1/2}\left(\frac{[q_\chi/m_\chi]}{10^{-2}} \right) \left( \frac{n_e}{10^{3}{~\rm cm}^{-3}} \right)^{1/2}  \left(\frac{V_{b\chi}}{40~{\rm km ~s^{-1}}}\right) ,
\end{eqnarray}
where we have evaluated at values most applicable for our CMB (see Table \ref{tab:values}) scenario described below (\S~\ref{subsec:CMBweibel}), the scenario that we find this warm-cold limit applies.  

\paragraph{Cold limit, where $\xi_j \gg 1$ for all participating species $j$:}
Finally, let us consider the case where all species are cold, the limit which yields the shortest instability times.  In this limit $Z(\xi_j) \approx -1/\xi_j$ + ${\cal O}(1/\xi_j)$ and the dispersion relation given by Equation \ref{eqn:fulldispersion} reduces to: 
\begin{equation}
0=c^2k^2 - \omega^2 + \omega_{\mathrm{p}p}^2 + \omega_{\mathrm{p}e}^2 + \omega_{\mathrm{p}\chi}^2+  f_\chi\frac{\omega_{\mathrm{p}\chi}^2}{2} \left( \frac{V_{b\chi} k}{\omega} \right)^2,
\end{equation}
which has solutions 
\begin{equation}
\omega = \frac{\pm 1}{\sqrt{2}} \left((\omega_{\mathrm{p}p}^2 + \omega_{\mathrm{p}e}^2 +    \omega_{\mathrm{p}\chi}^2 + k^2c^2) \pm \left[(\omega_{\mathrm{p}p}^2 + \omega_{\mathrm{p}e}^2 +    \omega_{\mathrm{p}\chi}^2 + k^2c^2)^2   + 2f_\chi\omega_{\mathrm{p}\chi}^2 V_{b\chi}^2 k^2 \right]^{1/2} \right)^{1/2}.\nonumber
\end{equation}

\noindent The positive, purely imaginary root is the classical Weibel solution, which has magnitude \citep{Weibel1959}:
\begin{equation}
\label{eqn:coldweibel}
    \gamma^{\rm C} \approx  \frac{  f_\chi^{1/2} \omega_{\mathrm{p}\chi} V_{b\chi} k}{\sqrt{2}\left(\omega_{\mathrm{p}j}^2  + c^2 k^2 \right)^{1/2}},
\end{equation}
which has a maximum of 
\begin{equation}
    \gamma_{\rm max}^{\rm C} \approx \omega_{\mathrm{p}\chi} f_\chi^{1/2}  \frac{V_{b\chi}}{c} =  2.8\times10^{-14}\rm{s}^{-1}\left( \frac{\rho_{DM}}{0.1 GeV cm^{-3}} \right)^{1/2} \bigg(\frac{[q_\chi/m_\chi]}{10^{-3}} \frac{V_{b\chi}}{4000~{\rm km~ s^{-1}}}\bigg),
\end{equation}
where the solution is valid for $|\xi_j| \gg 1$, which corresponds to wave numbers much greater than
\begin{equation}
    \label{eqn:kmax_coldweibel}
    \frac{\gamma^{\rm C}(k=k_{\mathrm{max}})}{k_{\mathrm{max}} \sigma_{T, j}} = 1 ~~\longrightarrow~~ ck_{\mathrm{max}} \approx \Bigg(\bigg(\frac{  f_\chi^{1/2} \omega_{\mathrm{p}\chi} V_{b\chi}}{\sqrt{2}\sigma_{T, j}}\bigg)^2 - \omega_{\mathrm{p}, j}^2\Bigg)^{1/2}.
\end{equation}
When $\omega_{\mathrm{p}, j}^2$ is the dark matter plasma frequency, a real $k_{\rm max}$ exists such that our solution holds. We find that this limit applies for the dark-$U(1)$ Bullet Cluster case (\S~\ref{subsec:BCweibel}). \\

\noindent We now detail how these growth rates apply to three astrophysical systems of interest.

\subsubsection{mDM streaming in the Early Universe}
\label{subsec:CMBweibel}

Dark matter is moving relative to baryons at the time of Recombination with an RMS velocity difference of $40\;$km s$^{-1}$  \citep{Hirata2010}. This Recombination-era dark matter cannot exchange significant momentum with the baryons or else it would disturb the percent level agreement between calculations in the $\Lambda$CDM model and the CMB.  Indeed, this observation has been used to constrain the DM millicharge under the approximation of collisional momentum exchange \citep[even if only $\sim 1\%$ of the DM is millicharged]{2014PhRvD..89b3519D, Boddy2018, 2021PhRvL.127k1301D}.

\begin{figure}[h!]
    \centering
    \includegraphics[scale=0.4]{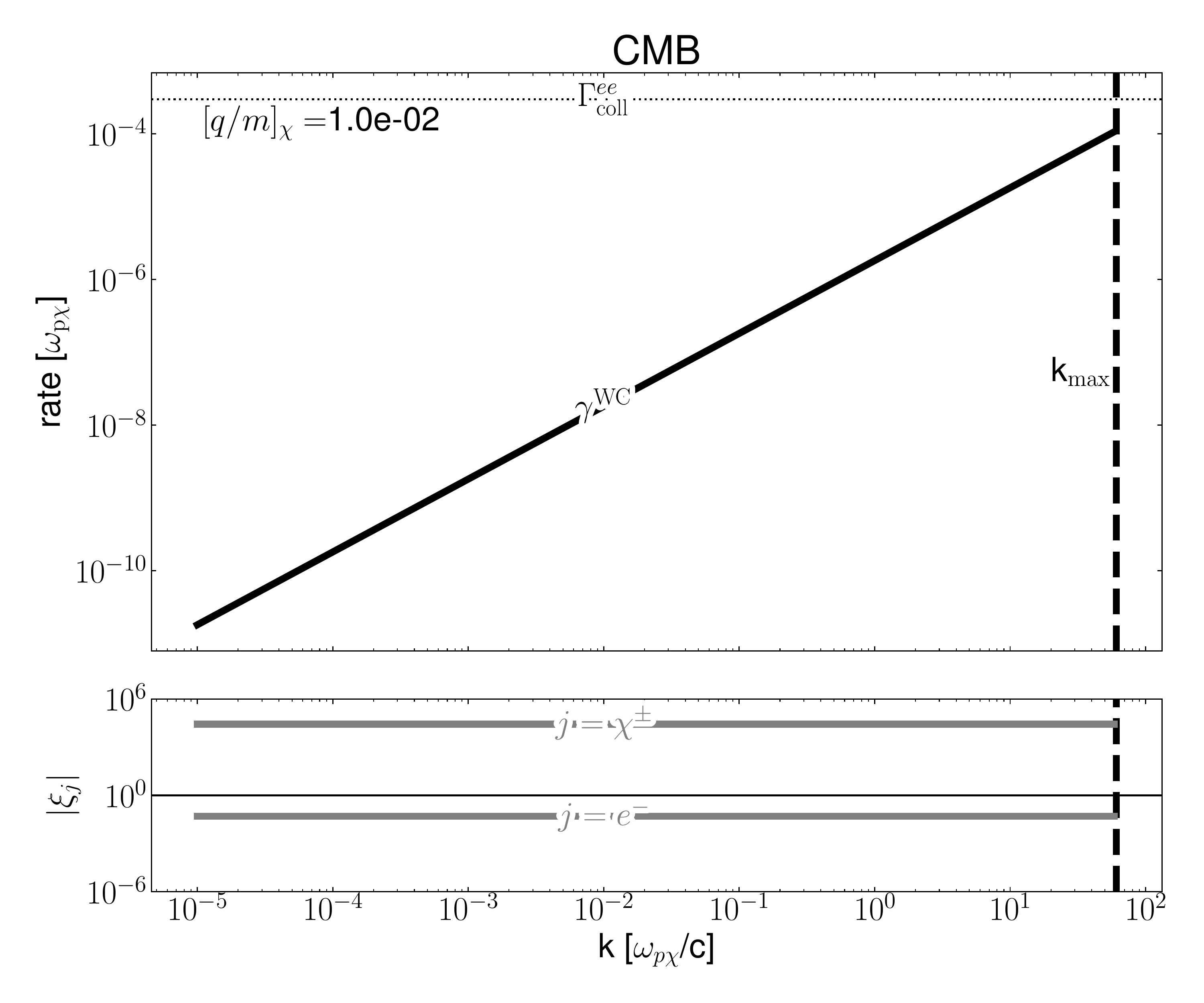}
     \caption{Millicharge DM instability growth and damping rates in pre-Recombination universe/CMB scenario calculated for $[q_{\chi}/m_{\chi}] = 10^{-2}$ such that $\omega_{\mathrm{p}\chi} = 7.9 \times 10^2$ Hz, with all other relevant fiducial parameters listed in Table~\ref{tab:values}.  {\it Top panel}: The black solid line show the warm-cold Weibel growth rate as a function of wavenumber(Equation~\ref{eqn:gammawarmcold}).  This analytic expression applies until approximately $k_{\rm max}$, indicated with the vertical dashed black line. The electron-electron collision rate $\Gamma_{\rm coll}^{ee} \sim 1$s$^{-1}$ is shown with the dotted black line.  At, $[q_{\chi}/m_{\chi}] = 10^{-2}$ the collision rate is always larger than the growth rate, damping the waves excited by the instability before the instability has time to grow.  Instabilities will only occur at somewhat larger values of $[q_{\chi}/m_{\chi}]$ than shown, values easily ruled out by other mDM bounds.  Thus, accounting for collective effects does not strengthen mDM constraints from CMB observations. {\it Bottom panel}: Demonstration that the warm-cold limit used in the top panel applies.  This limit holds when $|\xi_{\chi^\pm}| \gg 1$ and $|\xi_e| \ll 1$, which are both satisfied.} 
    \label{fig:CMBgamma_vsk}
\end{figure}

We find that plasma streaming instabilities are unlikely to improve on the CMB constraints from direct collisions. The ``CMB'' row in  Table~\ref{tab:values} shows the parameters used for our fiducial Recombination Era; the density of $10^3$cm$^{-3}$ corresponds to a redshift of $1700$. Figure~\ref{fig:CMBgamma_vsk} shows the growth rates for these parameters, evaluating the Weibel growth rate in the warm-cold limit (Equation~\ref{eqn:gammawarmcold}) for $[q_\chi/m_\chi] = 10^{-2}$ -- a charge-to-mass ratio easily ruled out by other constraints.\footnote{Unlike in the other physical situations we consider, for this CMB case we do not show a direct numerical evaluation of the full dispersion relation, Equation~\ref{eqn:fulldispersion}, however, we note that we were able to find numerical solutions near $k \approx k_{\rm max}$, verifying our analytic solution in this regime. } The bottom panel in Figure~\ref{fig:CMBgamma_vsk} shows that the warm-cold limit is applicable as $\xi_e \ll 1$ and $\xi_\chi \gg 1$.  Unfortunately, two-particle electron collisions damp the waves excited by the Weibel instability before the instability can grow, except for large charge-to-mass ratios that are ruled out with other methods. For our choice of $[q_\chi/m_\chi] = 10^{-2}$, the electron-electron collision rate is $\Gamma_{\rm coll}^{ee} \sim 1\;$s$^{-1}$ (shown by the dotted horizontal line).  The growth rate must be in excess of this damping rate in order for the instability to grow, but Figure~\ref{fig:CMBgamma_vsk} shows that is not the case for $[q_\chi/m_\chi] = 10^{-2}$: The growth rate is a factor of few below the collision rate at $k_{\rm max}$, the maximum wavenumber where our calculation for the warm-cold growth rate applies (see Equation \ref{eqn:gammawarmcoldmax} and the preceding text). We find that growth can happen faster than collisions at $k\sim k_{\rm max}$ only for $[q_\chi/m_\chi] \gtrsim 0.1$.  However, these large $[q_\chi/m_\chi]$ that would be ruled out by our instability analysis are already easily ruled out by studies that considered just collisions \citep{2014PhRvD..89b3519D, Boddy2018, 2021PhRvL.127k1301D}.

\begin{table}[tbp]
\centering
\begin{tabular}{|c|c|c|c|c|c|}
\hline
system & $V_{b, \chi}$ $[{\rm km~s^{-1}}]$ & $\sigma_{T, \chi}$ $[{\rm km~s^{-1}}]$ & $\sigma_{T, e}$ $[{\rm km~s^{-1}}]$  & $n_e$ $[{\rm cm}^{-3}]$ & $f_{\chi}$ \\
\hline
CMB & $4 \times 10^1$ & ${\rm negligible}$ & $4.6 \times  10^2$  & $1000$ & $1$  \\
MW & $4 \times 10^2$ & ${\rm variable}$  & $8.6 \times 10^3$   & $10^{-4}$ & $0.5$ \\
BC & $4 \times 10^3$ & $(1-3) \times 10^3$ & $4.3 \times 10^4$ & $2 \times 10^{-4}$ & $0.5$ \\
\hline
\end{tabular}
\caption{\label{tab:values} Fiducial parameters for the three astrophysical scenarios in which dark matter streaming can occur relative to baryons or other dark matter: The pre-Recombination universe as observed by the Cosmic Microwave Background (CMB; \S~\ref{subsec:CMBweibel}), the Milky-Way (MW; \S~\ref{subsec:MWweibel}), and the Bullet Cluster (BC; \S~\ref{subsec:BCweibel}). Here, $V_{b, \chi}$ is the relative velocity between the baryons and the DM, $\sigma_{T, \chi}$ is the thermal velocity of the DM,  $\sigma_{T, e}$ is the thermal velocity of the electrons, and $n_e$ is the electron number density of the systems.  We discuss how our results depend on these choices.}
\end{table}

\subsubsection{Bullet Cluster} \label{subsec:BCweibel}

The Bullet Cluster (1E 0657-56) has famously been used to place some of the strongest constraints on interacting dark matter models.  It consists of two galaxy clusters that are colliding, where the dark matter of the smaller cluster has passed through the larger (as indicated by lensing observations), whereas its gas has not passed through and is visibly shocking in the interior (as indicated by X-ray emission).  These clusters have velocity dispersions of $\approx 2000$ km s$^{-1}$ and are streaming through one another with velocity of $\approx 4000$ km s$^{-1}$ \citep{Markevitch2004}.  The $\approx 2000$ km s$^{-1}$ is from adding in quadrature and dividing by $\sqrt{2}$ the internal velocity dispersions of the two clusters, which we estimate to be $1500~$km s$^{-1}$ and $2200$ km s$^{-1}$ calculated from estimates of their temperatures from X-ray observations \citep{Markevitch2002} and the virial relation for ionized gas $\sigma_{T, \chi}^2 = 2k_b T/(0.59 \, m_p)$.  Given the crudeness of approximating the two-cluster system as having a single temperature in our calculations, we consider the instability for the range $1000-3000\,$km s$^{-1}$. If the Weibel instability can operate, it would have coupled the dark matter to the gas (or the dark matter to the dark matter in the case of a dark-$U(1)$ force) and would not be allowed to stream through itself, as is observed.  Thus, we can use this system to place constraints on mDM. We use the ``BC'' parameters in Table~\ref{tab:values} to evaluate the potential for instability. \\

\begin{figure}[h!]
    \centering
    \includegraphics[scale=0.4]{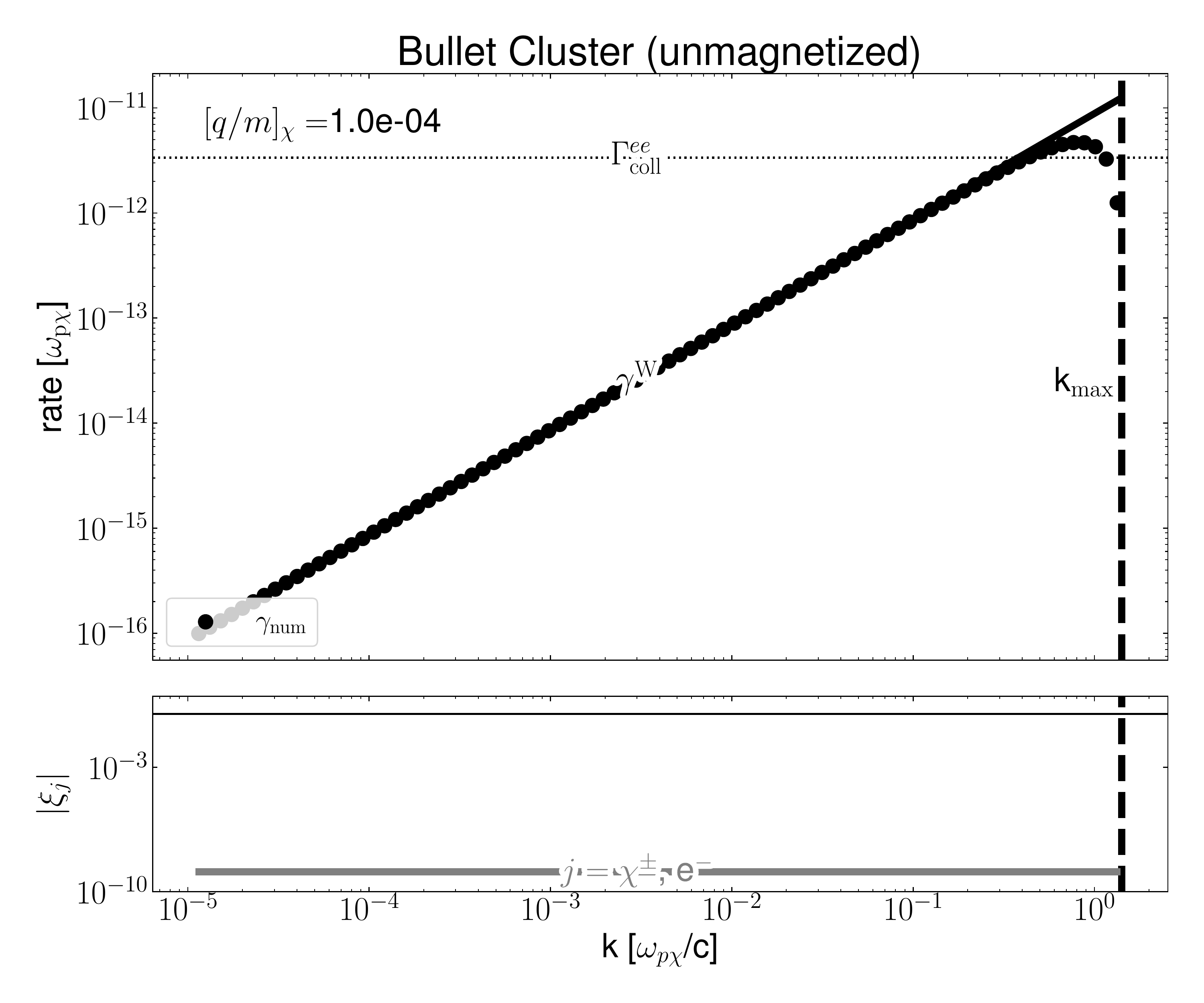}
    
    \caption{
    Millicharge DM instability growth and damping rates for the Bullet cluster, calculated for $[q_{\chi}/m_{\chi}] = 10^{-4}$ such that $\omega_{\mathrm{p}\chi} = 3.5\times10^{-3}$ Hz, and $\sigma_{T, \chi} = 2000$ km s$^{-1}$, with all other relevant fiducial parameters listed in Table~\ref{tab:values}. {\it Top panel}: The black solid line shows the warm-warm Weibel growth rate as a function of wavenumber (Equation~\ref{eqn:gammawarm}), the black points are the numerical solutions to the full dispersion given by Equation~\ref{eqn:fulldispersion}.  This analytic expression applies until approximately $k_{\rm max}$,   where we define $\gamma^{\rm W}_{\rm max}$ (Equation \ref{eqn:gammawarmmaxSM}), indicated with the vertical dashed black line. The electron-electron collision rate $\Gamma_{\rm coll}^{ee}$ is shown with the dotted black line.  At $[q_{\chi}/m_{\chi}] = 10^{-4}$ the collision rate is just slightly larger than the growth rate.  Instabilities will only occur at somewhat larger values of $[q_{\chi}/m_{\chi}]$ than shown, resulting in our Bullet Cluster constrain $[q_{\chi}/m_{\chi}] \gtrsim 10^{-4}$. {\it Bottom panel}: Demonstration that the warm-warm limit used in the top panel applies. The $|\xi_j| = 1$ horizontal line is shown in black for reference.  Indeed, the electrons and dark matter are warm with $|\xi_j| \sim 10^{-9} $ for the fiducial parameters (see gray curve).
    }
    \label{fig:bullet_vsk}
\end{figure}

\noindent {\bf mDM-induced instability:}
First, we consider the SM millicharge case, where the system parameters are in the limit of the Weibel instability for which both the SM particles and the dark matter are warm ($\xi_j \ll 1$). Using~Equation~\ref{eqn:gammawarm} and Equation~\ref{eqn:eecollisions}, we can solve for when the electron collision time is longer than the growth rate of the instability, indicating the instability can occur. We also solve the full Weibel dispersion (Equation \ref{eqn:fulldispersion}) using our numerical method detailed in Section \ref{sec:numericalmethod}, which agrees well with our derived analytic solution given by Equation \ref{eqn:gammawarm}. This is illustrated in Figure \ref{fig:bullet_vsk}.  {\it We find that $[q_\chi/m_\chi] \gtrsim 10^{-4}$ is excluded if the plasma is unmagnetized.}  However, galaxy clusters are known to have micro-Gauss large-scale magnetic fields \citep{govoni2004}.  In \S~\ref{sec:withmagneticfields} we show that this allows for modes that grow even  faster, resulting in even stronger limits.
\\

\begin{figure}[h!]
    \centering
    \includegraphics[scale=0.4]{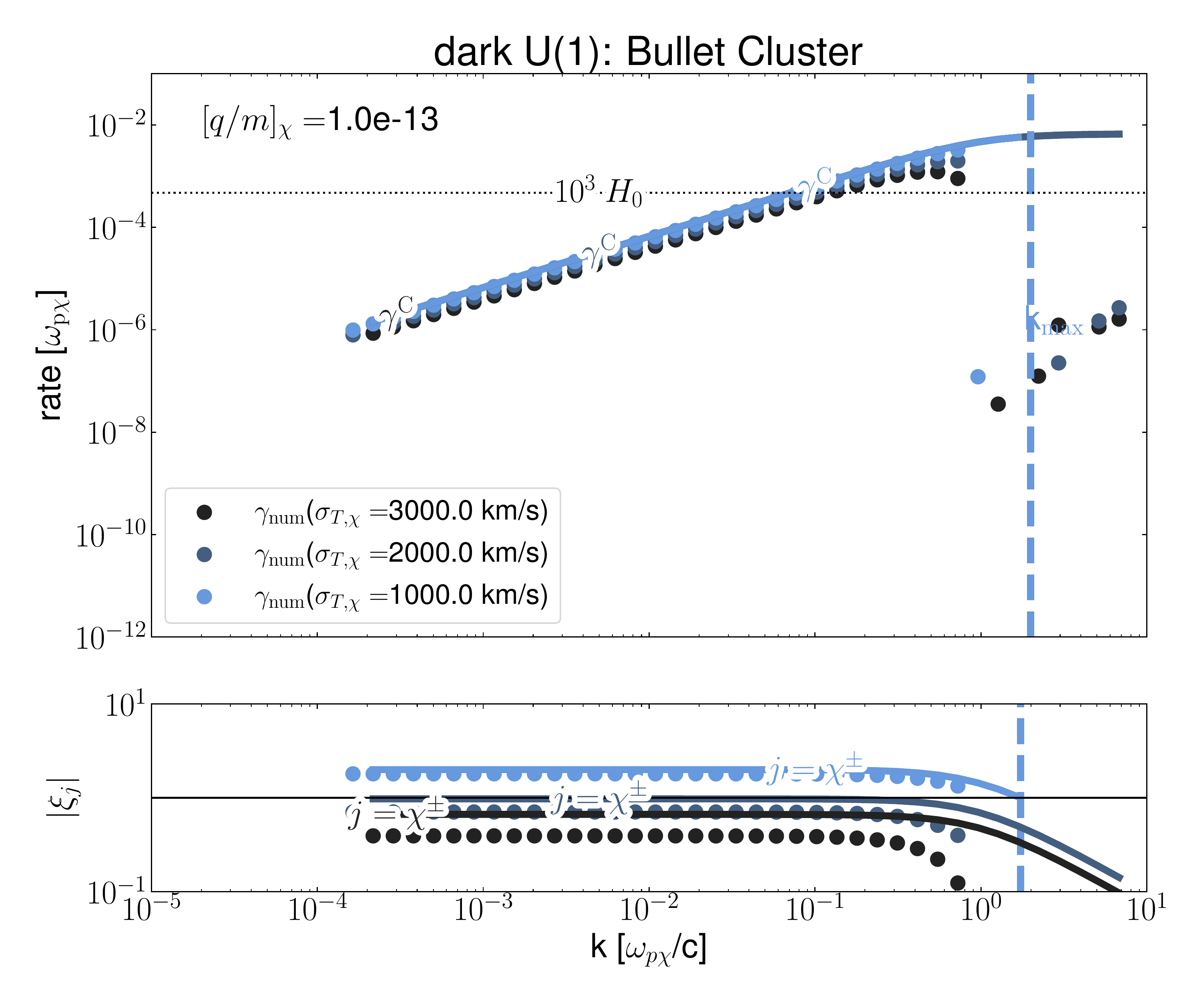}
    \caption{The Bullet Cluster system Weibel instability growth rates for dark matter charged under a dark-$U(1)$ with a massless dark photon. Calculations use $[q_{\chi}/m_{\chi}] = 10^{-13}$ and $\omega_{\mathrm{p}\chi} = 3.5 \times 10^{-12} $ Hz, considering our fiducial and more extreme velocity dispersion, $\sigma_{T,\chi}$, for the dark matter in the merging systems; the extreme values are chosen to show that our numerical solution slowly runs away from our analytic solution with increasing velocity dispersion. All other parameters used in the calculation are listed in Table~\ref{tab:values}. {\it Top panel}: The light blue, medium blue, and black points are the numerical solutions of the full Weibel dispersion relation Equation~\ref{eqn:fulldispersion} with the $\sigma_{T, \chi}$ = 1000, 2000 and 3000 km s$^{-1}$, respectively. The light blue, medium blue, and black solid lines show the analytic solution given by Equation~\ref{eqn:coldweibel} in the cold limit for the three respective $\sigma_{T,\chi}$. Our analytic approximation overpredicts the growth rate by $\approx 20\%$ for $\sigma_{T,\chi}= 1000$~km s$^{-1}$ up to $k \approx k_{\rm max}$ (Equation~\ref{eqn:kmax_coldweibel}; the dashed vertical lines are colored by the corresponding $\sigma_{T,\chi}$), where the cold  approximation used to derive our analytic solution no longer holds. In our fiducial case where $\sigma_{T, \chi}$ = 2000 km s$^{-1}$, we are no longer in the cold limit as $|\xi_{\chi}| \sim 1$ but we continue to find solutions that are within a factor of two of $\gamma^{\rm C}$. When $\sigma_{T, \chi}$ = 3000 km s$^{-1}$ (and for $\sigma_{T, \chi}$ > 2000 km s$^{-1}$), $|\xi_{\chi}| < 1$ and we find solutions that slowly depart from $\gamma^{\rm C}$.
    {\it Bottom panel}: Evaluation of $\xi_{\chi}$ at $\gamma^{\rm C}$ shown in light blue, medium blue, and black solid lines corresponding to $\sigma_{T, \chi}$ = 1000 km s$^{-1}$ 2000, and 3000 km s$^{-1}$, respectively. The light blue, medium blue, and black points correspond to $\xi_{\chi}$ evaluated at the numerical solution $\gamma^{\rm C}$ at a given $k$. This demonstrates the values of $\sigma_{T, \chi}$ that are in the cold limit and those which are not.
    }
    \label{fig:darkgamma_vsk}
\end{figure}

\noindent {\bf Dark-$U(1)$-induced instability:}
Next, we consider the case where there is a dark-$U(1)$ charge with a massless dark photon, but no SM millicharge.  This case turns out to be much more constraining, as the fast motions of electrons don't damp the instability as in the SM millicharge case. {\it In particular, we find  as long as $[q_\chi/m_\chi] \gtrsim 10^{-14}$, that there are always modes that are able to grow and would reach a nonlinear amplitude in the age of the system.}  This is shown in Figure~\ref{fig:darkgamma_vsk}, which considers a charge-to-mass of $[q_\chi/m_\chi] =10^{-13}$, near our claimed constraint.  The horizontal dotted line is $10^3 H_0$, the rate required to grow $100$ $e$-foldings in the cluster crossing time, taken to be $ 0.1(H_0)^{-1}$ or approximately a billion years. (The size of the system divided by the infall velocity yields a similar timescale.)  The light blue, medium blue, and black solid lines are our analytic cold-limit estimates for $\sigma_{T,\chi} = 1000$, 2000, and 3000 km s$^{-1}$, respectively.  

The light blue, medium blue, and black points correspond to the respective velocities and are numerical solutions to the full dispersion relation. We find these solutions using the method detailed in Section \ref{sec:numericalmethod}, with the initial simplex for the numerical method started around  $\omega_0 = \gamma^{\rm C}$, our derived analytic solution given in Equation~\ref{eqn:coldweibel}. In our fiducial case where $\sigma_{T, \chi}$ = 2000 km s$^{-1}$, we are at the borderline of the cold limit with $|\xi_{\chi}| \sim 1$.  For smaller values of $\sigma_{T, \chi}$, we are able to reproduce the cold limit, finding numerical solutions that are within $\sim 20\%$ of $\gamma^{\rm C}$ until $k \approx k_{\rm max}$ (given by Equation \ref{eqn:kmax_coldweibel}), above which the cold limit, $|\xi_j| \gg 1$, no longer holds.  This is illustrated by the  $\sigma_{T, \chi}$ = 1000 km s$^{-1}$ points in Figure~\ref{fig:darkgamma_vsk}, which fall on the corresponding theory curve below $k_{\rm max}$. In the fiducial case where $\sigma_{T, \chi}$ = 2000 km s$^{-1}$, we are no longer in the cold limit and $|\xi_{\chi}| \sim1$ (see bottom panel of Figure \ref{fig:darkgamma_vsk}), but we still find numerical solutions which are within a factor of two of $\gamma^{\rm C}$. As we push to larger values of $\sigma _{T, \chi} > 2000$ km s$^{-1}$, where $|\xi_{\chi}| < 1$,  our numerical solution slowly moves away from $\gamma^{\rm C}$.  This shows that the growth rate is not very sensitive to the assumed cold limit and, hence, our parameter choices.  This is in contrast to the two-stream (longitudinal) instability, which shuts off for $\sigma _{T, \chi}/V_{b,\chi} >1/\sqrt{3} \approx 0.6$ \citep{1997aspp.book.....T}.  The Bullet cluster system is likely above this threshold.

\subsubsection{the Milky Way system} 
\label{subsec:MWweibel}

We briefly comment on the likelihood of instability in the Milky Way halo and other galactic-mass dark matter halos. In falling dark matter is likely more unstable to Weibel than in the Bullet cluster system just considered because the velocity dispersion of the satellites or unvirialzed DM falling onto the Milky Way can be considerably smaller relative to the in fall velocity, making the DM effective colder and, hence, more unstable. Thus, for the unmagnetized case, we expect instability happens in the Milky Way in the SM mDM case for $[q_\chi/m_\chi] > 10^{-4}$ and for the dark-$U(1)$ charge for $[q_\chi/m_\chi] > 10^{-13}$.  The constraints on the former case are even much stronger when we include magnetic fields, as discussed in \S~\ref{sec:withmagneticfields}, and Milky Way-like dark matter halos may harbor $\sim 1\mu$G magnetic fields \citep{2021MNRAS.501.4888V}.

 If instabilities occur in the Milky Way or other galactic mass halos, we expect the dark matter to behave more like a gas.  Behaving like a gas may reduce the triaxiality of the dark matter halo, and it will also affect the strong constraint on mDM from disk spin-down.  Constraints from these considerations are discussed in \S~\ref{sec:previousconstraints}.

\subsection{Weibel Instability with non-zero background magnetic field}
\label{sec:withmagneticfields}
We now turn to the case where there is a non-zero background magnetic field, $B_0$. The presence of the magnetic field  introduces magnetic degrees of freedom and timescales (set by the Larmor frequencies of charged species, see Table \ref{tab:plasmaparams}). Beyond this, magnetic fields make plasmas look colder perpendicular to $B_0$, as in their presence, charged particles become confined; this should help Weibel instabilities to grow on shorter time scales at a given charge-to-mass ratio.  Indeed, \cite{LiLin2020} showed that mDM-baryon streaming instabilities in supernovae shocks can be hugely amplified in the presence of magnetic fields.  We note that while we consider the presence of magnetic fields in the mDM case, we do not consider them in the dark-$U(1)$ case, as there is no strong motivation for large scale magnetic fields there. 

The magnetized Weibel instability has a homogenous field that is perpendicular to the direction of propagation.  The presence of a perpendicular field is a bit counterintuitive for the problem at hand, as such a large-scale field would not allow charged particles to stream. However, when the two systems are streaming through each other (the most applicable situation being two clusters falling into each other), sometimes the field will be parallel and sometimes perpendicular.  In the perpendicular case, without instability it is likely the streaming will exert force on the perpendicular field lines that aligns them with the direction of streaming, as there is so much energy in the streaming dark matter.  The presence of the Weibel instability would not allow this realignment of the large-scale field, as it excites small scale fields that couple the fluids and likely magnetic turbulence.  The other case, where the field is aligned with the streaming motion, is discussed in the \S~\ref{sec:firehose}, where we find similar growth rates to here.

With this setup, let us now consider the mathematics in various limits. In the cold limit for all species $\xi_j \gg 1$ (and ignoring helium and metals), we can expand the Weibel plasma dispersion relation, Equation \ref{eqn:fulldispersion}:
\begin{equation}
\label{eqn:mag_coldweibel}
\begin{split}
     0 = c^2 k^2 - \omega^2 + \omega_{\mathrm{p}p}^2 \left(\frac{\omega}{\omega \pm \Omega_p} \right) + \omega_{\mathrm{p}e}^2 \left(\frac{\omega}{\omega \pm \Omega_e} \right) \\
    +  \omega_{\mathrm{p}\chi}^2 \left(\frac{\omega^2}{\omega^2 - \Omega_\chi^2} \right) +   f_\chi \omega_{\mathrm{p}\chi}^2 V_{b\chi}^2 k^2 \frac{\omega^2 + \Omega_\chi^2}{(\omega^2 - \Omega_\chi^2)^2},
\end{split}
\end{equation}
The cold or marginally cold limit when $\xi \gtrsim 1$ is the only interesting limit as the effect of magnetic fields is to make the plasma appear colder and enhance instability. 
Let us additionally take the limit where  $\omega \ll \Omega_b$, as we anticipate Weibel-like solutions with small growth rates, such that the sum over the baryonic species term reduces to: 
 \begin{equation}
 \label{eqn:highmag_baryons}
 \begin{split}
     \sum_{b=i^+,e^-} \omega_{\mathrm{p}b}^2 \bigg( \frac{\omega}{\omega \pm \Omega_b}\bigg)  \approx \sum_{b= i^+, e^-} \frac{\omega_{\mathrm{p}, b}^2\omega}{\Omega_b} \bigg(1-\omega/\Omega_b\bigg),
     ~~~~~~~~~~~~~~~~~~~~~~~~~~~~~~~~~\\ 
     = \omega\bigg[\frac{\omega_{\mathrm{p}, e}^2}{\Omega_e}\bigg(1-\frac{\omega}{\Omega_e}\bigg) + \frac{\omega_{\mathrm{p}, i}^2}{\Omega_{i}}\bigg( 1- \frac{\omega}{\Omega_i}\bigg)\bigg] = -\omega^2c^2  \bigg(\frac{\sqrt{4\pi(\rho_e+\rho_i)}}{B_0}\bigg)^2 = -\frac{\omega^2c^2}{v_{A}^2},
 \end{split}
 \end{equation}
 where we have used that $q_{e^-} = - q_{i^+}$ and ${\omega_{\mathrm{p}, j}^2}/{\Omega_j} = 4\pi n_j q_j c/{B_0}$, where $B_0$ is the homogeneous background magnetic field oriented along the direction of the perturbation and orthogonally to the streaming motion.  Equation \ref{eqn:mag_coldweibel} becomes: 
\begin{equation}
\begin{split}
     0 = c^2 k^2 - \omega^2 \bigg(1 + \frac{c^2}{v_A^2} \bigg)
    +  \omega_{\mathrm{p}\chi}^2 \left(\frac{\omega^2}{\omega^2 - \Omega_\chi^2} \right) +   f_\chi \omega_{\mathrm{p}\chi}^2 V_{b\chi}^2 k^2 \frac{\omega^2 + \Omega_\chi^2}{(\omega^2 - \Omega_\chi^2)^2}.
\end{split}
\label{eqn:beforeotherlimits}
\end{equation}

We now consider two limits of this equation.  First, in the very low frequency limit where $\omega \ll{} \Omega_{\chi}$, the previous equation becomes:
\begin{equation}
\label{eqn:highmag_coldweibel}
\begin{split}
     0 = c^2 k^2 - \omega^2 \bigg(1 + \frac{c^2}{v_A^2} \bigg)
    - \omega_{\mathrm{p}\chi}^2 \left(\frac{\omega}{\Omega_{\chi}} \right)^2 +  \frac{f_\chi \omega_{\mathrm{p}\chi}^2 V_{b\chi}^2 k^2 }{\Omega_{\chi}^2} \bigg(1 + 3\frac{\omega^2}
    {\Omega_{\chi}^2}\bigg),
\end{split}
\end{equation}
which has solution
\begin{equation}
\label{eqn:highmag_coldweibelsoln}
\begin{split}
     \omega^2 = \bigg(c^2 +  \frac{f_{\chi} \omega_{\mathrm{p}\chi}^2V_{b\chi}^2}{\Omega_{\chi}^2}\bigg)k^2 / \bigg(1 + \frac{c^2}{v_A^2} + \frac{\omega_{\mathrm{p}\chi}^2}{\Omega_{\chi}^2} -  \frac{3f_\chi \omega_{\mathrm{p}\chi}^2 V_{b\chi}^2 k^2 }{\Omega_{\chi}^4} \bigg).
\end{split}
\end{equation}
\noindent We can derive a criterion on the maximum wavenumber where our solution holds by requiring $\xi_j = 1 $:
\begin{equation}
\label{eqn:highmag_coldweibel_kmax}
\begin{split}
     k_{\rm max}^2 = \bigg[\bigg(1 + \frac{c^2}{v_A^2} + \frac{\omega_{\mathrm{p}\chi}^2}{\Omega_{\chi}^2}\bigg) -  \bigg(\frac{c^2}{\sigma_{T, j}^2} +  \frac{f_{\chi} \omega_{\mathrm{p}\chi}^2V_{b\chi}^2}{\Omega_{\chi}^2\sigma_{T, j}^2}\bigg)\bigg] / \bigg(\frac{3f_\chi \omega_{\mathrm{p}\chi}^2 V_{b\chi}^2 }{\Omega_{\chi}^4} \bigg).
\end{split}
\end{equation}
\noindent Unfortunately, there are no modes which yield imaginary solutions to Equation \ref{eqn:highmag_coldweibelsoln} and simultaneously obey Equation~\ref{eqn:highmag_coldweibel_kmax}.  We conclude that this very low frequency limit does not yield an interesting solution.

Next, we consider the opposite higher frequency limit where $\Omega_{\chi} \ll \omega$.  In this limit Equation~\ref{eqn:beforeotherlimits} becomes
\begin{equation}
\label{eqn:lowmag_coldweibel}
\begin{split}
     0 = c^2 k^2 - \omega^2 \bigg(1 + \frac{c^2}{v_A^2} \bigg)
    - \omega_{\mathrm{p}\chi}^2  +  \frac{f_\chi \omega_{\mathrm{p}\chi}^2 V_{b\chi}^2 k^2 }{\omega^2},
\end{split}
\end{equation}
\noindent which has solutions: 
\begin{equation}
\begin{split}
     \omega^2 = \Bigg((c^2k^2 - \omega_{\mathrm{p}, \chi}^2) \pm \sqrt{(c^2k^2 - \omega_{\mathrm{p}, \chi}^2)^2+4f_{\chi}\omega_{\mathrm{p}, \chi}^2 V_{\mathrm{b},\chi}^2k^2\bigg(1 + \frac{c^2}{v_A^2}\bigg)}\Bigg) / \bigg(2 (1 + \frac{c^2}{v_A^2})\bigg),
\end{split}
\end{equation}
\noindent which again yields the ordinary cold Weibel growth rate given by Equation~\ref{eqn:coldweibel} in the limit that the second term under the square root is small compared to the $c^2k^2-\omega_{{\rm p}, \chi}^2$ term and for the purely imaginary root.  Thus, this limit also does not yield an interesting solution.

Since we were unable to find interesting solutions in the previous limits, we consider the marginally cold case where $\omega \sim i \Omega_{\chi}$.  Unfortunately, in this case a similar analytic approach is not possible, and instead we are forced to search blindly for numerical solutions.

\begin{figure}[h!]
    \centering \hspace*{-0.33cm}
    \includegraphics[scale=0.28]{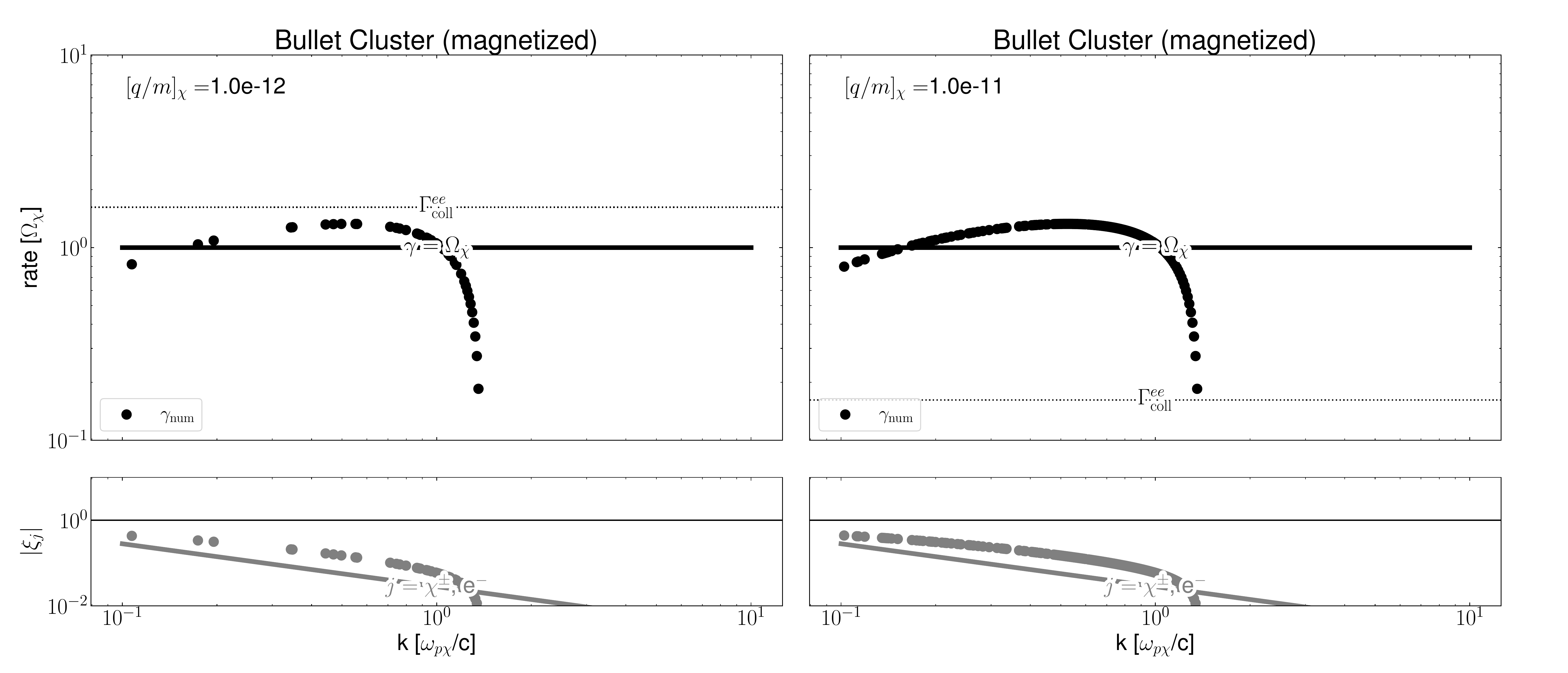} 
    \caption{Numerical evaluation of Weibel dispersion for the Bullet Cluster system for the parameters in Table~\ref{tab:values}, at $\sigma_{T, \chi} = 2000$ km s$^{-1}$, and assuming a background magnetic field with $B_0 = 1 \mu $G as motivated by cluster observations. The left panel assumes $[q_{\chi}/m_{\chi}] = 10^{-12}$ such that $\omega_{\mathrm{p}\chi} = 3.5 \times 10^{-11}$Hz and $\Omega_{\chi} = 9.6 \times 10^{-15}$Hz, and the right panel $[q_{\chi}/m_{\chi}] =10^{-11}$ such that $\omega_{\mathrm{p}\chi} = 3.5 \times 10^{-10} $ Hz and $\Omega_{\chi}= 9.6 \times 10^{-14}$ Hz.  {\it Top panels}: The black, solid line corresponds to our initial guess frequency, $\omega_0 = i\Omega_{\chi}$, when solving the magnetized Bullet Cluster dispersion case numerically. The numerical solutions are relatively stable when  wavenumber is between $0.1-10 [\omega_{{\rm p}, \chi}/$c] and peak with maximum growth rates of $\gamma_{\rm max} \sim i\Omega_{\chi}$. When $[q_{\chi}/m_{\chi}] = 10^{-12}$, $\Gamma^{ee}_{\rm coll}$ is larger than $\gamma_{\rm max}$, and the Weibel instability is unable to grow. However, $\gamma_{\rm max}>\Gamma^{ee}_{\rm coll}$ for $[q_{\chi}/m_{\chi}] = 10^{-11}$, which sets our lower bound on the charge-to-mass ratio of $\chi$ for the magnetized BC.  This lower bound is approximately seven orders of magnitude lower than the unmagnetized Bullet Cluster bound we set in \S~\ref{subsec:BCweibel}. {\it Bottom panels}: Evaluation of $\xi_j$ at $\gamma_{\rm num}$ for all participating species $j$. Since we find $|\xi_j| \lesssim 1$, the analytic approach taken elsewhere would not work to find these solutions.  
    }  
    \label{fig:B0BC_rate_vsK}
\end{figure}

Of the situations we consider, the presence of a large-scale magnetic field is most motivated for the Bullet Cluster, as one does not expect the Recombination era gas to be strongly  magnetized. We use $B_0 = 1\mu$G, as observations show that cluster magnetic fields are at the $\mu$G level \cite{govoni2004}.
We initialize our numerical method as detailed in Section \ref{sec:numericalmethod} around the order-of-magnitude guess $\omega_0 = i\Omega_{\chi}$.  This guess is motivated by the mDM-baryon streaming instability growth rates found in \cite{LiLin2020} in the case of supernovae shocks.  However, fortuitously, this guess does find interesting solutions:  We find solutions with growth rates exceeding the electron-electron collision rate down to $[q_{\chi}/m_{\chi}] = 10^{-11} $ (right upper panel of Figure \ref{fig:B0BC_rate_vsK}).  However, when  $[q_{\chi}/m_{\chi}] \leq 10^{-12} $  (upper left panel of Figure \ref{fig:B0BC_rate_vsK}), the electron-electron collision rate exceeds the growth rate of the instability such that collisions will damp the instability. {\it Thus, we find $[q_\chi/m_\chi] \gtrsim 10^{-12}(B_0/\mu)$G
is excluded if the plasma is magnetized}, seven orders of magnitude lower than when the mDM-plasma is unmagnetized as detailed in Section \ref{subsec:BCweibel}.

\subsection{Firehose Instability} \label{sec:firehose}
In the case where the mDM propagation direction is parallel to the background magnetic field, ${\bf B}_0$ (i.e. ${\bf V}_{\rm{b}, \chi} \parallel {\bf B}_0$), we expect the enhancement of EM perturbations to give way to a beam-fire hose instability.  Physically, the firehose instability is induced by a back reaction against the centrifugal force as charged baryons and dark matter move along curving magnetic lines of force.  The result of the instability is likely plasma turbulence, with the beam coupled to the background plasma. 

A small perturbative field ${\bf \delta B}$ redirects charged particles with a Lorentz force and produces a drift current in the  direction of $\nabla \times {\bf B}$ \citep{Parker1958}. In this case, the dispersion relation is given by
\begin{equation}
\label{eqn:firehose}
    0 = D^{\pm}(k, \omega) = c^2k^2 - \omega^2 - \sum_{b = i^+, e^-} \omega_{\mathrm{p}b}^2 \bigg( \frac{\omega}{k\sigma_{T, b}}\bigg) Z(\xi_b) - \sum_{s = \chi^+, \chi^-} \omega_{\mathrm{p}, s}^2 \bigg(\frac{ \omega - k V_{b, \chi}}{k\sigma_{T, \chi}}\bigg) Z(\xi_s)
 \end{equation}
where $\xi_b = (\omega\pm\Omega_b)/k\sigma_{T, b}$ and $\xi_s = (\omega-kV_{b, \chi}\pm\Omega_s)/k\sigma_{T, \chi}$ and $Z(\xi_j)$ is the plasma dispersion function defined in Appendix~\ref{ap:dispersionfunction}.

We again consider the Bullet Cluster with a background magnetic field of ${ B}_0 = 1 \mu$G. In this case we considered only numerical solutions with $\gamma \approx i \Omega_{\chi}$ as in \citep{LiLin2020}. Indeed, solving Equation \ref{eqn:firehose} numerically, we find numerical solutions with $\gamma_{\rm max} \approx i \Omega_{\chi}$ over orders of magnitude in $[q_{\chi}/m_{\chi}]$ down to $[q_{\chi}/m_{\chi}] \approx 10^{-11}$. Below this, we find electron-electron collisions, given by Equation~\ref{eqn:eecollisions} damp waves excited by the instability.

\section{Effects of plasma instabilities on previous constraints} \label{sec:previousconstraints}
Some of the strongest constraints on mDM arise from the interaction with magnetic fields in galaxies, possibly spinning down the galactic disk \citep{Stebbins2019}, and magnetic fields in clusters, potentially resulting in different halo profiles \citep{2016arXiv160204009K}.  The plasma instabilities we have considered excite strong small-scale magnetic fields, which may alter these constraints.  We briefly comment on these constraints in light of our results.

\subsection{Halo Profile Constraints} Ambient magnetic fields could change the radial density profiles of galaxy clusters for the case of mDM. Indeed, \cite{2016arXiv160204009K} claims the strongest constraints on mDM by requiring the mDM gyro radius for observed cluster ambient magnetic fields of $\sim 1\,\mu G$ exceed the $\sim 1$ Mpc size of a galaxy cluster. 

Dark matter-baryon plasma instabilities, excited as dark matter falls onto the galaxy, would increase the dark matter baryon coupling and perhaps strengthen any deviations from the collisionless-limit NFW halo profiles. For example, if the instability was induced, the ambient small-scale magnetic field strength could increase causing the gyro radius, $r_{L\chi}$, to decrease further. Then, to maintain the inequality $r_{L\chi} \gtrsim 1$ Mpc from \citep{2016arXiv160204009K}, the derived bound on $[q_\chi/m_\chi]$ would decrease. On the other hand, it is unclear to us that even the hydrodynamic limit (small gyro radius) is ruled out, as hydrodynamic-only simulations produce similar profile halos to NFW \citep[e.g.][]{2017MNRAS.470..500N}.  So, we avoid deriving constraints from halo radial profiles.

Another interesting constraint owes to the observed asphericity of galaxy clusters.
Simulations show that galaxy clusters should be more spherical if the dark matter is interacting, which may be in conflict with observations \citep{Peter2013}. Even in the hydrodynamic limit of many interactions, the timescale for a sound wave to cross the virialized system and restore isotropy is still the dynamical time, the same timescale for a collisionless system.   Thus, this constraint relies on the ${\cal O}(1)$ difference in the exact anisotropy-damping timescale for an interacting and truly collisionless system. Whether constraints apply to saturating plasma instabilities, which for example may not be exactly the hydrodynamic limit, is unclear.  Additionally, for mDM, anisotropy could be further enhanced by astrophysical feedback, which will have more of an effect on the dark matter since the dark matter is more coupled to the gas.  This enhanced coupling could perhaps even play a role in the coring of dwarf galaxy halos and the cusp-core problem \citep[e.g][]{Bertoni2014}.

\subsection{Disk spin-down constraints} \label{subsec:ISMspindown} It was argued in \citep{Stebbins2019} that as mDM passes through the interstellar medium (ISM) disk of spiral galaxies, the mDM will be deflected by embedded ordered magnetic fields. In particular, it is argued that as mDM passes through the disk of a spiral galaxy, it will be deflected by the magnetic fields and thus angular momentum will be exchanged between the rotating ISM and the slowly rotating mDM halo. This exchange would then cause the entire ISM disk to spiral inwards, in contrast with observations, allowing for constraints to be placed on the charge-to-mass ratio of mDM. Stebbins \citep{Stebbins2019} considered the case of ballistic trajectories into the Milky Way disk, deriving the tight bound
\begin{equation}
\frac{q_X}{e} \lesssim 10^{-12}\left(\frac{m_X}{m_p} \right) \left( \frac{v}{300 {\rm ~km~s^{-1}}} \right)\left(\frac{\mu G}{B} \right)\left( \frac{\rm kpc}{\ell_B} \right),
\end{equation}
where $\ell_B$ is coherence length of galactic magnetic field. This bound is roughly set by the dark matter having a Larmor radius comparable to the width of the Galactic disk for a $\sim \mu$G Galactic magnetic field.

If plasma instabilities couple the dark matter to the gas within halos, the natural saturation of such instabilities is likely with the magnetic field coming into kinetic energy equipartition with the gas, which in the Milky Way halo results in $B \sim \sqrt{6\pi n k T_{\rm vir}}\sim  1\mu G$.  (Equipartition magnetic fields may also be generated by astrophysical processes.)  The dark matter then diffuses through the resulting magnetic inhomogeneities, rather than traveling ballistically through the Galactic disk as assumed in \cite{Stebbins2019}.  The Weibel instabilities excites small scale magnetic inhomogeneities, below the Larmor radius of the dark matter.  The later nonlinear stages of the instability likely lead to turbulence and a very inhomogeneous magnetic field, and we assume there is some component that points in different directions on the scale of the dark matter Larmor radius. In this circumstance, a simple picture for how the dark matter travels takes that the dark matter scatters in a different direction each time it traverses its gyro-radius leading to the estimate $D_\chi \sim v_\chi r_{L\chi}$ for the diffusion coefficient, where $r_{L\chi} = m_\chi c v_\chi /[q_\chi B]$ is the DM gyro-radius. Plugging in values yields
\begin{equation}
D = v_\chi r_{L\chi}
= 1.4\times 10^{-2} {\rm ~kpc}^2 \; {\rm Gyr}^{-1} \left( \frac{v_X}{200 {\rm ~km~s^{-1}}} \right)^2 \left(\frac{[q_\chi/m_\chi]}{10^{-8}}\right)^{-1} \left( \frac{B}{1~\mu G}\right)^{-1}.
\end{equation}

The mechanism of \cite{Stebbins2019} uses that the ISM baryons in the galactic disk transfers its angular momentum to the halo.  In their case, they considered ballistic trajectories for which dark matter travels from far out in the halo into the disk.  Instead, especially if instabilities like those considered in this paper occur (although a tangled magnetic field from astrophysical processes can do the same), the dark matter diffuses through the halo, exchanging momentum between adjacent regions and bringing them into solid body rotation.  For substantial spin-down of the disk, we need $\sim 10\%$ of the dark matter to come into solid body rotation with the disk, since $\sim 10$ percent of the mass in our halo is in the disk.  Very roughly, this requires the inner $R\sim 30~$kpc to come into solid body rotation (i.e. assuming a $r^{-2}$ profile that extends out to $300$kpc), which occurs over a timescale of
\begin{equation}
\tau_{\rm solid body} \sim  \frac{R^2}{D} = 70 \; {\rm Gyr}\; \left(\frac{R}{30~ {\rm kpc}}\right)^2 \left(\frac{v_\chi}{200 {\rm ~km~s^{-1}}}\right)^{-2}\left(\frac{[q_\chi/m_\chi]}{10^{-11}}\right) \left( \frac{B}{1 ~\mu G}\right)
\end{equation}
Equating this with the $\sim 10~$Gyr age of the Milky Way and evaluating at our fiducial values of $R \sim 30 ~{\rm kpc}$ and $v_{\chi} \sim 200$ km s$^{-1}$, {\it millicharges with $[q_{\chi}/m_{\chi}] \gtrsim 10^{-12}$  may not be ruled out by disk spin-down}.  We conclude that much of the parameter space ruled out by \cite{Stebbins2019} could be affected by halo magnetic fields leading to the dark matter taking more diffusive trajectories and, hence, slowing spin-down.  

\section{Discussion} \label{sec:discussion} 

\begin{figure}[h!]
    \centering \hspace*{-1.2cm}
    \includegraphics[scale=0.5]{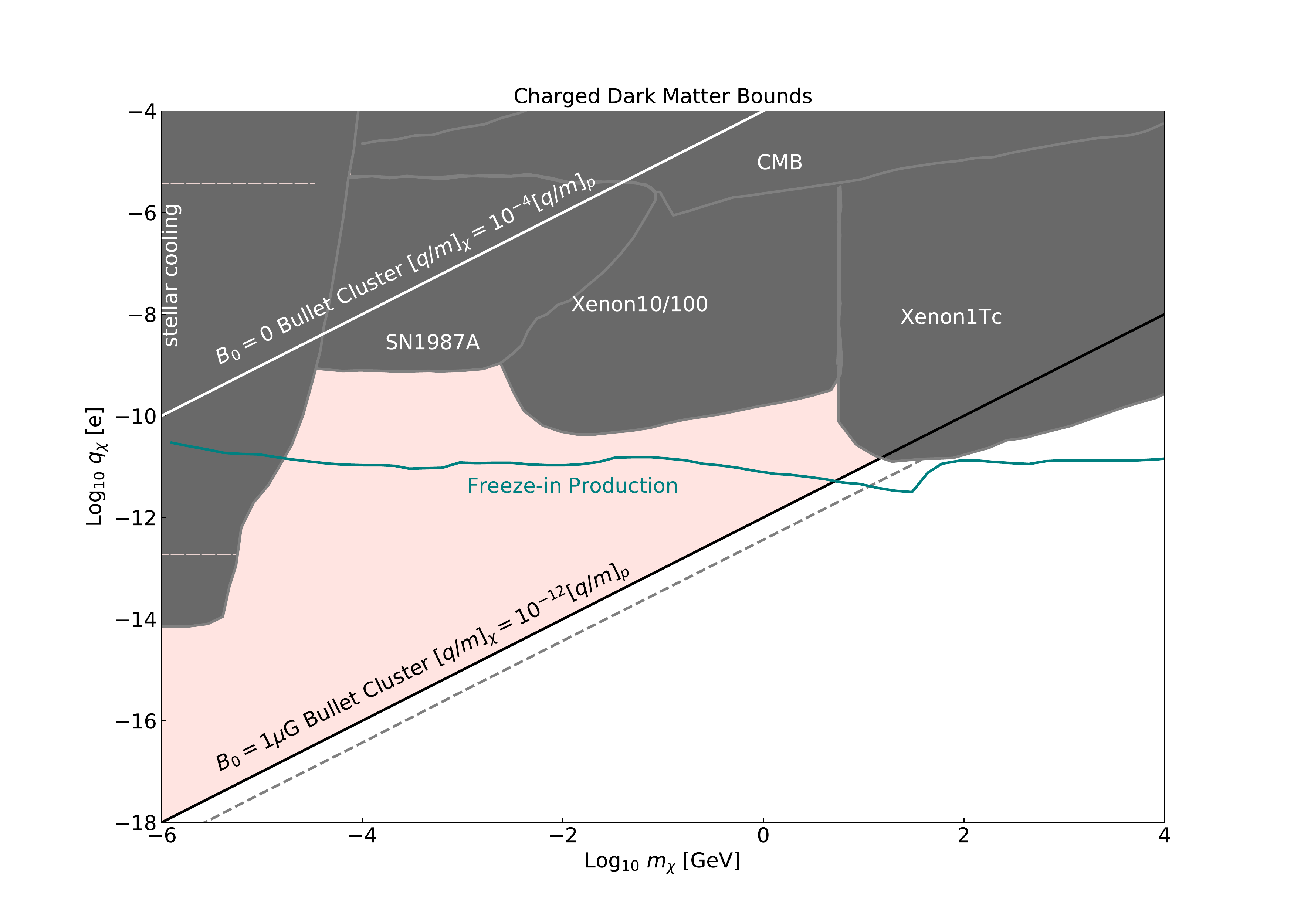}
    \caption{mDM bounds in the $q_{\chi}$--$m_{\chi}$ plane. Above the white and black solid lines, the mDM charge-to-mass ratio $[q_{\chi}/m_{\chi}]$ is sufficient to drive Weibel instabilities in the Bullet Cluster when the magnetic field is 0 and 1 $\mu$G, respectively.  Since 1 $\mu$G is inline with observations, we consider the latter much more stringent constraint to be our most realistic and, hence, the pink shaded region indicates our exclusion region. Above the upper dashed, gray line, Stebbins \cite{Stebbins2019} argues $[q_{\chi}/m_{\chi}]$ is robustly ruled out as the mDM would spin-down the Milky Way ISM. In Section \ref{subsec:ISMspindown}, we conclude that when $[q_{\chi}/m_{\chi}] \gtrsim 10^{-12}$, mDM could be affected by halo magnetic fields, leading to mDM taking diffusive, rather than ballistic trajectories as assumed in \cite{Stebbins2019}, pushing their constraints below the black solid line to the gray dashed line. The solid teal line represents the parameter space for freeze-in production of mDM, where annihilation of SM particles gives rise to the abundance after inflation and before matter-radiation equality \citep{Chu2012, Dvorkin2019}. The gray shaded region shows previous constraints on the existence of mDM by the SLAC mDM experiment \citep{Prinz1998}, stellar cooling \citep{Vogel2014}, and  supernova 1987a \citep{Chang2018}. Additional constraints are shown from CMB decoupling \citep{Kovetz2018, McDermott2011}, and electron and nuclear recoil direct detection in Xenon10/100 \citep{Essig2012, Essig2017} and Xenon1Tc, respectively.}
    \label{fig:mDM_bounds}
\end{figure}


\begin{figure}[h!]
    \centering \hspace*{-1.2cm}
    \includegraphics[scale=0.5]{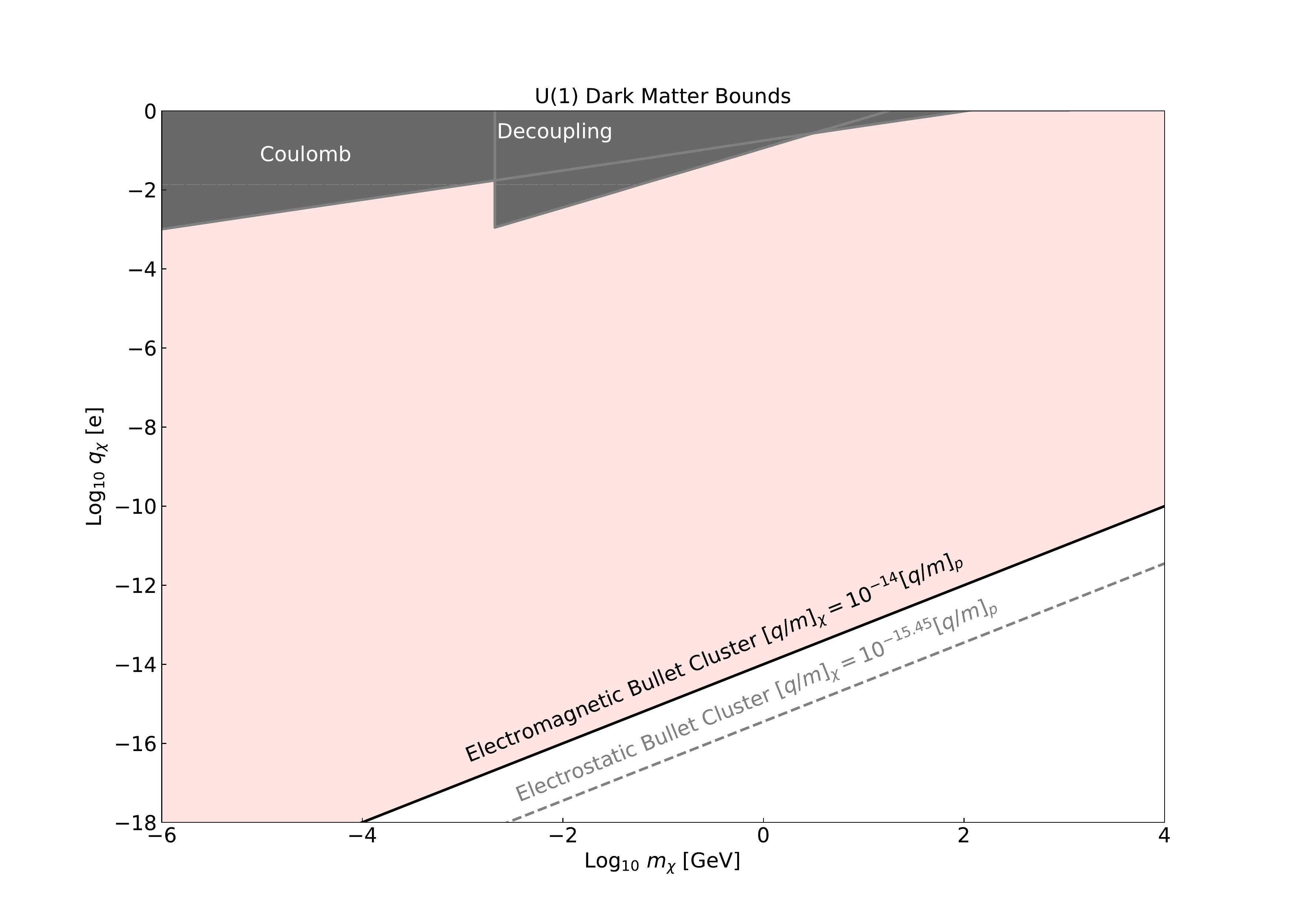}
    \caption{Bounds  in the $q_{\chi}$--$m_{\chi}$ plane on dark matter that is $U(1)$ charged with a massless dark photon. In the pink shaded region above the solid black electromagnetic line, the dark-$U(1)$ DM charge-to-mass ratio is sufficient to drive Weibel instabilities in the Bullet cluster system as detailed in section \ref{sec:weibel}.  This instability was considered previously in \cite{Lasenby2020} in the cold limit and we showed that it still grows sufficiently fast in the warm limit.  Also shown is the electrostatic instability growth rate \citep{Lasenby2020}, which can grow faster by a factor of c/$V_{\rm{b},\chi}$, but we argue is a less robust constraint in the Bullet cluster.  The gray shaded Coulomb region represents the parameter space where 2 $\rightarrow$ 2 Coulomb collisions would have a significant impact in DM halos as detailed in \citep{Lasenby2020}. The solid gray shaded region is disfavored by kinetic decoupling constraints detailed in \citep{Heikinheimo2015}. While our solutions considered a massless dark photon, we note that these results should hold for a massive photon whose Compton wavelength is larger than the length scale of the instability, $\sim c/\omega_{p \chi}$.
    }
    \label{fig:U1_bounds}
\end{figure}

Our main findings are summarized in Figures \ref{fig:mDM_bounds} and \ref{fig:U1_bounds} along with constraints derived in previous works. Starting with mDM constraints in Figures \ref{fig:mDM_bounds}, the solid teal line in Figure \ref{fig:mDM_bounds} shows the required mDM charges and masses for freeze-in production of all the dark matter.  Freeze-in is a scenario where the DM is created from the byproducts of annihilation of SM particles after inflation and before matter-radiation equality \citep{Chu2012, Dvorkin2019}. 
The dark gray regions in Figure \ref{fig:mDM_bounds} represent collected bounds on mDM production. Bounds from a dedicated beam dump preformed at SLAC are shown for charges $q_{\chi} > 10^{-5} q_e$ \citep{Prinz1998}. Stellar energy losses due to the emission of mDM pairs by plasmon decay would have reduced the neutrino pulse of SN1987A, owing to constraints derived in \citep{Chang2018}. Low mass constraints from 
anomalous cooling from the emission of mDM from white dwarves, red giants, and horizontal branch stars are shown and labeled as stellar cooling \citep{Vogel2014}. Additional constraints are shown from CMB decoupling \citep{Kovetz2018, McDermott2011}, and electron and nuclear recoil direct detection in Xenon10/100 \citep{Essig2012, Essig2017} and Xenon1Tc, respectively.  Above the gray dashed line, Stebbins \cite{Stebbins2019} argues $[q_{\chi}/m_{\chi}]$ is robustly constrained by ballistic mDM, ISM spin-down. However, in Section \ref{subsec:ISMspindown}, we conclude that when $[q_{\chi}/m_{\chi}] \gtrsim 10^{-12}$, mDM could be affected by halo magnetic fields, leading to mDM taking diffusive, rather than ballistic, trajectories as assumed in \cite{Stebbins2019}, pushing the constraints from Milky Way spin-down to the band between the dashed gray line and the solid black line. Above the white solid line, we show our weaker unmagnetized-Weibel constraint derived earlier in this work (\S~\ref{subsec:BCweibel}) in the absence of a magnetic field. Our more realistic, magnetized exclusion region from electromagnetic streaming instabilities derived are indicated by the pink shaded region above black solid lines, where the mDM charge-to-mass ratio $[q_{\chi}/m_{\chi}]$ is sufficient to drive Weibel instabilities in the Bullet Cluster when the magnetic field is 1 $\mu$G, consistent with observations (\S~\ref{sec:withmagneticfields}).

Now turning to dark-$U(1)$ constraints detailed in Figure \ref{fig:U1_bounds}. The gray shaded regions show constraints from $2 \rightarrow 2$ Coulomb collisions that would significantly impact DM halos \citep{Lasenby2020}. The shaded region labeled ``decoupling'' show constraints from \citep{Heikinheimo2015} derived by considering kinetic decoupling. Specifically, the authors obtain a temperature of kinetic decoupling, $T_{\rm kin}$, by equating the Hubble rate to the Compton scattering rate for the dark plasma. They then require that decoupling happens above $T_{\rm kin} > 640$ eV, so that DM-dark radiation coupling only significantly influences CMB multipoles above $l > 2500$. The pink shaded region above the solid black Electromagnetic Bullet Cluster line represents our derived constraint on the dark-$U(1)$ DM charge-to-mass ratio that is sufficient to drive unmagnetized Weibel instabilities in the Bullet cluster system. These constraints corroborate the earlier bounds of \citep{Lasenby2020, Ackerman2009}, and further serve to confirm these results by showing that the growth rate of the instability in the cold limit (Equation \ref{eqn:coldweibel}) is a reasonable approximation for the numerical parameters consistent with the Bullet Cluster (e.g. values listed in Table \ref{tab:values}).  This is even though the Bullet cluster systems parameters is only at the borderline of the cold limit.  We further showed that even for higher velocity dispersions than seem reasonable for this system, the growth rates do not change so significantly, indicating the robustness of the linear Weibel instability to the assumed parameters. 

However, the strongest constraints on dark-$U(1)$ DM come if electrostatic instabilities are able to couple the counter-streaming momenta. Electrostatic instabilities grow faster than the electromagnetic instabilities by a factor of $c/V_{b, \chi}$, or $\sim 1000$ in the Bullet cluster system.  An estimate for the parameter space ruled out by electrostatic instabilities assuming $100$ efoldings charge-to-mass ratios is given by the black line shown in Figure \ref{fig:U1_bounds} \citep{Lasenby2020}.  However, electrostatic instabilities, like the two-stream, generally saturate by flattening the bump on the tail of the distribution.  Such saturation may occur before substantial momentum exchange \citep[e.g.][]{1997aspp.book.....T}. In addition, such instabilities will not grow at all if $\sigma _{T, \chi}/V_{b,\chi} >0.57$ \citep{1997aspp.book.....T}, where the Bullet cluster system is near equality.  Thus, the electromagnetic Weibel instability considered here may be more robust. 

Finally, we briefly discuss the general case when the dark matter has both a dark and millicharge.  The dispersion relations are worked out for this possibility in Appendix \ref{ap:darkandmilli}. We find that for the purely dark Weibel dispersion relation to apply, as we had considered in \S~\ref{subsec:BCweibel}, the ratio of the dark matter's dark charge to its SM charge, $Q$, must satisfy $Q \gg 1$. Thus, our dark Bullet Cluster constraint is only valid if the dark-$U(1)$ charge is much greater than the SM millicharge. On the other hand, our mDM constraints are only valid when the dark charge is either zero or very small, such that $Q\ll 1$.  Zero is most likely case since mDM models where the millicharge is created by a slight mixing with a dark charge should have a significant $Q$. See Appendix \ref{ap:darkandmilli} for more discussion.


\section{Conclusions} \label{sec:conclusion}
We derived new and corroborated previous bounds on mDM and dark-$U(1)$ DM with an extremely light dark photon by considering plasma streaming instabilities. These instabilities act to couple the momenta of counter-streaming plasmas, likely resulting in the dark matter becoming unable to penetrate either baryonic plasma or another dark matter plasma.  Such coupling would conflict with cosmological situations where the dark matter is observed to stream through the baryons or other dark matter.  Landau damping likely highly damps electrostatic instabilities when the DM is millicharged, owing to the fact that the streaming velocity dispersion of particles is comparable to their streaming motion for the cosmological situations where such counter-streaming may occur. When DM instead has a dark-$U(1)$ charge, electrostatic instabilities are not prohibited by Landau damping and grow faster than electromagnetic instabilities \cite{Lasenby2020}, although electromagnetic instabilities may be more likely to couple the counter-streaming plasmas' momenta. 
Thus, we focused on electromagnetic instabilities. 
 
 We considered three physical situations where dark matter is streaming relative to itself or other dark matter.  
 \begin{enumerate}
     \item  The first is the cosmic microwave background, where the baryons and dark matter are moving relative to each other at Recombination.    However, we found that electron two-particle collisions suppress streaming instabilities in this situation.  Thus, the streaming of dark matter in the pre-Recombination universe is affected more strongly by direct two-particle collisions rather than collective processes, validating previous constraints.
     \item The second situation was the Bullet Cluster, where the dark matter in one cluster is streaming through the dark matter and baryons in the other.  First, considering the instructive case of the Weibel instability in an unmagnetized plasma with mDM, the properties of the Bullet Cluster merger and other merging cluster systems would result in fast enough growth of the linear instability to likely couple the momenta if $[q_\chi/m_\chi] \gtrsim 10^{-4}$.  For smaller $[q_\chi/m_\chi]$, electron collisions damp the waves excited by the instability.   When a homogeneous magnetic field is considered with amplitude consistent with cluster observations, both the Weibel and Firehose instabilities have much larger growth rates of approximately the cyclotron frequency owing to the new magnetic waves that can be excited.  This more realistic magnetized case results in our final constraint being $[q_\chi/m_\chi] \gtrsim 10^{-12}$. 
     \item   Finally, we considered a dark-$U(1)$ charge and the mutual coupling of the counter-streaming dark matter.  We find that the Weibel instability is likely fast enough to grow to saturation in the age of the Bullet cluster system for $[q_\chi/m_\chi] \gtrsim 10^{-14}$, in agreement with \cite{Lasenby2020}.  Our work extended the study of Lasenby \cite{Lasenby2020} and Ackerman \citep{Ackerman2009} by verifying that the cold approximation for electromagnetic Weibel growth still roughly holds despite the parameters of this system being only at the borderline of this limit. We found that as we pushed to increasingly larger values of $\sigma _{T, \chi}$ above $2000$ km s$^{-1}$ where $|\xi_{\chi}| < 1$ such that the cold limit does not hold, our numerical solution only slowly moved away from the cold growth rate, which indicates the robustness of the linear Weibel instability. While our solutions considered a massless dark photon, our results should hold for a massive photon whose Compton wavelength is larger than the length scale of the instability, $\sim c/\omega_{p \chi}$.
 \end{enumerate}

 


The final physical situation we considered is the Milky Way and its surrounding halo. We showed that the previous constraints from Milky Way ISM spin-down of \cite{Stebbins2019} are weakened in a scenario where plasma instabilities, or even just tangled background magnetic fields, couple mDM to the gas within a dark matter halo. If such coupling happens, the dark matters propagation is likely to instead be diffusive, instead of traveling on ballistic trajectories, making spin-down timescales longer than the age of the Milky Way unless $[q_{\chi}/ m_{\chi}] \gtrsim 10^{-12}$.  We further discussed whether other observations of dark matter halos may rule out such diffusive propagation.

We conclude with one caveat to our analysis. We assumed that if the linear growth rate of a considered instability is below the collision rate that damps its excited waves, then the instability grows until it saturates with a significant enough magnetic field to deflect particles.  For the Weibel instability, many studies have found that it saturates when it grows to large enough amplitudes to deflect the particles participating on the small scale of the excited modes, and such a saturation mechanism makes physical sense in the context of this purely growing instability.  However, since our study considers scenarios far from the baryonic case that has been studied in detail, it is possible that nonlinear parasitic processes are able to damp these slowly growing instabilities before they reach large enough magnetic fields for deflection.  More work is required to address this possibility.

\section{Acknowledgments}

AC was partially supported by the National Science Foundation Graduate Research Fellowship under Grant No. DGE-1762114. 
The authors would like to thank Robert Lasenby for comments on an early draft of this work. 
\appendix

\section{Plasma dispersion function and expansions}
\label{ap:dispersionfunction}
For Maxwellian distributions functions (and ``drifting'' Maxwellians), it is useful to define the \textit{plasma dispersion function}, 

\begin{equation} \label{eq:plasma_dis_func}
    Z(\xi_j) \equiv \frac{1}{\sqrt{\pi}}\int_{-\infty}^{\infty} \frac{e^{-x^2}}{x - \xi_j}
\end{equation}
The plasma dispersion function is the Hilbert transform of a Gaussian. It is useful to express this function in its limiting cases:  
\begin{equation}
    Z(\xi_j) = i\sqrt{\pi}\mathrm{exp}(-\xi_j^2) - 2\xi_j+\frac{4}{3}\xi_j^3 - \frac{8}{15}\xi_j^5 + \dots, \hspace{0.5cm} \mathrm{ \hspace{0.5cm} 
    for \hspace{0.2cm} } |\xi_j| < 1
\end{equation}
and
\begin{equation}
    Z(\xi_j) = i\sigma\sqrt{\pi}\mathrm{exp}(-\xi_j^2) - \frac{1}{\xi_j} - \frac{1}{2\xi_j^3} - \frac{3}{4\xi_j^5} + \dots, \hspace{0.5cm} \mathrm{ \hspace{0.5cm} 
    for \hspace{0.2cm} } |\xi_j| > 1
\end{equation}

\noindent where $\xi_j = x + iy$ and 

\begin{equation}
    \sigma = \left\{ 
    \begin{array}{ccc}
        0  &  & y > 1/|x|;  \\
        1  &  & |y| < 1/|x|; \\ 
        2  &  &  y < -1/|x|.
    \end{array} \nonumber
    \right.
\end{equation}
These expansions are used in evaluating the limiting growth rates.

\section{Dark matter with dark and millicharge} \label{ap:darkandmilli}
When the dark matter has both a dark and millicharge, the dispersion relation for the right and left circular electromagnetic modes is even more complicated but can be calculated from the determinant of the dielectric tensor to find modes that satisfy $D_{ij} E_j =0$, where $\vec E_{\rm all} = (E^{+}, E^-, {\cal E}^+, {\cal E}^-)$ where ${\cal E}$ is the dark electric field in the left and right circular polarization basis and

\begin{equation}
\mathbf{D} =
    \begin{pmatrix} 
    \Tilde{\omega}^2 - k^2 (S^+_{\rm SM} + S^+_{\chi}) & 0 & - k^2 Q S^+_{\chi} & 0\\
                     0 &  \Tilde{\omega}^2 - k^2 (S^-_{\rm SM} + S^-_{\chi}) & 0 & - k^2 Q S^-_{\chi}\\
                     - k^2 Q S^+_{\chi} &  0  & \Tilde{\omega}^2 -  k^2 Q^2 S^+_{\chi} & 0 \\
                     0& - k^2 Q S^-_{\chi} & 0 &  \Tilde{\omega}^2 -  k^2 Q^2 S^-_{\chi}
    \end{pmatrix}
\end{equation}

\noindent where $\Tilde{\omega}^2 = c^2k^2 - \omega^2$.  To satisfy our equation $D_{ij} E_j =0$ for some $E_j$ requires $0 = \det( \mathbf{D})$, which reduces to the dispersion relation
\begin{equation}
 0 = \left[c^2 k^2 - \omega^2 - k^2 (S^\pm_{\rm SM} + S^\pm_{\chi}) \right] \left[c^2 k^2 - \omega^2 - k^2 Q^2 S^\pm_{\chi} \right] -  k^4 Q^2 (S^\pm_{\chi})^2,
 \label{eqn:darkdispersion}
\end{equation}
where the $S^{\pm}$ can be identified from our dispersion relations for the Weibel and Firehose cases (Equations~\ref{eqn:fulldispersion} and \ref{eqn:firehose}), as these previous dispersion relations are derived from the determinant of the upper $2\times2$ quadrant of $\mathbf{D}$, and the $Q$ is the ratio of the dark matter's dark charge to its SM charge.  Note that the dark electric field terms have the same polarization for the $\xi$ as the particle paths are still shaped by the background SM magnetic field.  (If the dark matter has no SM charge, then $\Omega_\chi = 0$ and this appendix is not relevant.) Equation~(\ref{eqn:darkdispersion}) has the familiar form of the dispersion relation we used for the millicharged case, times that for the dark-$U(1)$ case, but then minus a cross term $k^2 Q^2 S^-_{\chi}$.  If we can neglect this cross term, then we can safely consider the limit of millicharged or dark instabilities, as done in the main paper.  This of course can be done for $Q=0$ but also the limit of $Q =\infty$.\footnote{To take the $Q =\infty$ limit, note $Q^n S^\pm_{\chi}$ goes to zero for $n<2$ and is finite for $n=2$ such that Equation~\ref{eqn:darkdispersion} reduces to multiplying two independent dispersion relations, one for the dark matter and the other for the baryons.}

For the Weibel $S^{\pm}$, the dispersion relation becomes
\begin{eqnarray} 
        0 &=&  \Bigg(c^2 k^2 - \omega^2 - \sum_{b = i^+, e^-} \omega_{p b}^2 \big(\frac{\omega}{k \sigma_{T, b}}\big) Z(\xi_{b}^{\pm})
        -  \sum_{s = \chi^+, \chi^-} (\omega^{\rm SM}_{ps})^2 \bigg[ \bigg(\frac{\omega}{k \sigma_{T, s}}\bigg)Z(\xi_{s}^{\pm}) \\
        &+& f_\chi \bigg(\frac{V_{b\chi}}{\sigma_{T, s}}\bigg)^2 (1 + \xi_{s}^{\pm} Z(\xi_{s}^{\pm})) \bigg] \Bigg) \Bigg(c^2 k^2 - \omega^2   \nonumber \\
        &-& \sum_{s = \chi^+, \chi^-} (\omega^{\rm dark}_{ps})^2 \bigg[ \bigg(\frac{\omega}{k \sigma_{T, s}}\bigg)Z(\xi_{s}^{\pm}) + f_\chi \bigg(\frac{V_{b\chi}}{\sigma_{T, s}}\bigg)^2 (1 + \xi_{s}^{\pm} Z(\xi_{s}^{\pm})) \bigg] \Bigg) \nonumber \\
        &-& \left(\sum_{s = \chi^+, \chi^-} \omega^{\rm dark}_{ps}\omega^{\rm SM}_{ps} \bigg[ \bigg(\frac{\omega}{k \sigma_{T, s}}\bigg)Z(\xi_{s}^{\pm}) + f_\chi \bigg(\frac{V_{b\chi}}{\sigma_{T, s}}\bigg)^2 (1 + \xi_{s}^{\pm} Z(\xi_{s}^{\pm})) \bigg] \right)^2 \nonumber,
\label{eqn:dispersionWithTwo Charges}
\end{eqnarray}
where we now distinguish between the dark and SM plasma frequencies for the dark matter with superscripts `SM' and `dark'.  This has the form of the dispersion relationship we solved for the millicharged case times the dispersion relation we assumed for the dark case, minus a cross term.  

For a purely dark instability to hold we require $\gamma_{\rm max}^2 \gg [k^2 S_\chi]_{\rm max}$ where `max' indicates that we are checking at the fastest growing mode where $k \sim \omega_{p\chi}/c$ for our unmagnetized Weibel, such that the zero is approximated by the solution to the dark case. This reduces to the intuitive condition $Q \gg 1$ for the unmagnetized cold Weibel. Thus, only when the dark charge is much greater than the SM millicharge is our dark Bullet Cluster constraint valid. Similarly, $\gamma_{\rm max}^2 \gg Q^2 [k^2  S_\chi]_{\rm max}$ for the SM electromagnetic instability.  This condition can be used to put limits on what dark charges are allowed to not damp our mDM instability, the most relevant case being the magnetized Weibel.  While we do not investigate this condition in detail, our mDM constraints are only valid with the dark charge is either zero or very small such that $Q\ll 1$.  Since zero is most likely since mDM models where the millicharge is created by a slight mixing with a dark charge generate a significant $Q$, we consider our mDM constraints only valid for dark matter with pure SM charge.





\bibliographystyle{JHEP.bst}
\bibliography{References}












\end{document}